\documentclass[preprint,12pt]{elsarticle}

\usepackage{amssymb}

\usepackage{amsmath}
\usepackage{amsfonts}
\usepackage{makeidx}
\usepackage{graphicx}
\usepackage{tensor}
\usepackage{color}
\usepackage{breqn}
\usepackage{graphicx}
\usepackage[colorlinks=True]{hyperref}
\usepackage[labelfont=bf]{caption}
\usepackage{subcaption}
\usepackage{float}
\usepackage{multirow}
\usepackage{multicol}
\usepackage{pifont}

\usepackage{tikz}
\newcommand{\tikzxmark}{%
\tikz[scale=0.23] {
    \draw[line width=0.7,line cap=round] (0,0) to [bend left=6] (1,1);
    \draw[line width=0.7,line cap=round] (0.2,0.95) to [bend right=3] (0.8,0.05);
}}
\newcommand{\tikzcmark}{%
\tikz[scale=0.23] {
    \draw[line width=0.7,line cap=round] (0.25,0) to [bend left=10] (1,1);
   \draw[line width=0.8,line cap=round] (0,0.35) to [bend right=1] (0.23,0);
}}

\begin{document}
\begin{frontmatter}



\title{\textbf{Sequential learning based PINNs to overcome temporal domain complexities in unsteady flow past flapping wings}}


\author[a]{Rahul Sundar}
\author[b]{Didier Lucor}
\author[a]{Sunetra Sarkar}

\address[a]{Department of Aerospace Engineering, Indian Institute of Technology Madras,Chennai, 600036 , Tamil Nadu, India}
\address[b]{Laboratoire Interdisciplinaire des Sciences du Numérique LISN-CNRS, Orsay, 91403, France}

\begin{abstract}
For a data-driven and physics combined modelling of unsteady flow systems with moving immersed boundaries, 
Sundar {\it et al.} \cite{sundar2024physics} introduced an immersed boundary-aware (IBA) framework, combining Physics-Informed Neural Networks (PINNs) and the immersed boundary method (IBM). This approach was beneficial because it avoided case-specific transformations to a body-attached reference frame. Building on this, we now address the challenges of long time integration in velocity reconstruction and pressure recovery by extending this IBA framework with sequential learning strategies. Key difficulties for PINNs in long time integration include temporal sparsity, long temporal domains and rich spectral content. To tackle these, a moving boundary-enabled PINN is developed, proposing two sequential learning strategies: - a time marching with gradual increase in time domain size, however, this approach struggles with error accumulation over long time domains; and - a time decomposition  which divides the temporal domain into smaller segments, combined with transfer learning it effectively reduces error propagation and computational complexity. The key findings for modelling of  incompressible unsteady flows past a flapping airfoil include: - for quasi-periodic flows, the time decomposition approach with preferential spatio-temporal sampling  improves accuracy and efficiency for pressure recovery and aerodynamic load reconstruction, and, - for long time domains, decomposing it into smaller temporal segments and employing multiple sub-networks,  simplifies the problem  ensuring stability and reduced network sizes. This study highlights the limitations of traditional PINNs for long time integration of flow-structure interaction problems and demonstrates the benefits of decomposition-based strategies for addressing error accumulation, computational cost, and complex dynamics.
\end{abstract}




\end{frontmatter}


\section{Introduction} \label{sec:Intro}
 Physics-informed neural networks~\cite{lagaris1998artificial,raissi2019physics} are a hybrid class of supervised and unsupervised learning models capable of learning from labeled data while encoding prior physical knowledge into its loss functions. 
 This has allowed PINNs to solve complex forward and inverse problems in fluid dynamics and other domains.
  Their framework might also be particularly well-suited in a context of multi-query problems, such as optimization,  data assimilation and uncertainty quantification \cite{Cheng_IEEE_2023}, and parametric exploration, as they provide a flexible and efficient framework for solving complex time-dependent systems across varying inputs without the need for repeated computationally expensive simulations.
 However, standard PINNs~\cite{raissi2019physics} could be  difficult to train due to various failure modes such as competing objectives~\cite{wang2020understanding}, spectral bias~\cite{wang2021eigenvector}, and propagation failures~\cite{krishnapriyan2021characterizing}, especially in the case of forward problems. 
Many authors have proposed adaptive loss component weighting~\cite{wang2020understanding, heydari2019softadapt}, domain decomposition strategies~\cite{jagtap2020conservative,dwivedi2021distributed}, adaptive activation functions~\cite{jagtap2020adaptive}, and modified architectures~\cite{wang2020understanding},  to mitigate the above failure modes. 
Systems with moving boundaries are more difficult for PINNs to capture because the domain evolves over time, complicating the enforcement of boundary conditions and increasing computational complexity. Additionally, moving boundaries often involve coupled nonlinear dynamics and can introduce sharp gradients or discontinuities in the solution, which are challenging for neural networks to approximate accurately.
In this context, some previous works relied on domain transformations to body-attached frame of reference~\cite{raissi2018deepVIV}. But this can be limiting when handling multiple moving bodies or flexible structures.
Hence, in the context of unsteady flows past moving/flapping wing-like bodies, the use of a fixed background Eulerian grid would be enviable to train PINNs. This served as the motivation for some of the recent works ~\cite{huang2022direct,yang2021fdm, calicchia2023reconstructing, sundar2024physics, zhu2023pinnsdyninterface, zhu2024physics}.

Huang {\it et al.}~\cite{huang2022direct} proposed a direct forcing immersed boundary method based PINN (IB-PINN) which was validated for a canonical steady-state flow past a fixed cylinder. Here, IB-PINN used the modified NS equations which included a forcing term in the momentum conservation equation. While the IB-PINN performance was discussed, a comparison with a PINN formulation considering the standard Navier-Stokes equations was missing. Recently, Calicchia {\it et al.}~\cite{calicchia2023reconstructing} used PINNs based on standard Navier-stokes equations to recover pressure from planar velocity field data obtained from PIV experiments for flow past swimming fish. In addition to the physics loss, the authors used data losses to fit the PIV velocity measurement data in the interior bulk, and a no penetration boundary condition loss at the moving interface. 
The results were promising, however, a deeper understanding of the relative merits of  standard NS-based PINNs or IBM-based PINNs was missing. 

Towards this, Sundar {\it et al.}~\cite{sundar2024physics} proposed an immersed boundary-aware framework for a moving boundary enabled PINN, that included a standard NS equations based formulation (MB-PINN) and an IBM based formulation (MB-IBM-PINN). Sundar {\it et al.}~\cite{sundar2024physics} showed MB-PINNs to be useful in scenarios where body position and velocity are known, and MB-IBM-PINN having potential in scenarios where such information might not be available {\it apriori}. Additionally, a physics-based vorticity-cutoff based sampling technique was proposed to improve the data sample efficiency of the models while maintaining good accuracy levels. In a connected work~\cite{sundar2023understanding}, using a novel zonal loss gradient splitting technique, it was shown how vorticity-cutoff based sampling technique alleviated vanishing gradients. 
More recently, Zhu {\it et al.}~\cite{zhu2024physics} demonstrated the capability of NS-based PINNs (same as MB-PINNs) in modeling forward and hidden physics recovery problems for flow past multiple flexible cylinders with predefined trajectories in three-dimensional spatial domains. 

Forward problems are not attempted in the current study given their well known difficulties in training PINNs over a long time domain in an error free manner~\cite{krishnapriyan2021characterizing, wang2022respecting, zhu2024physics}. These difficulties get further accentuated for moving body problems~\cite{sundar2024physics, zhu2024physics}. This is especially true when the dynamics strongly depends on the accuracy of the flow-field time history that is being resolved. 
For forward problems, traditional CFD solvers always triumph over PINNs, which are yet to achieve a similar level of proficiency. However, PINNs work very well for inverse problems such as hidden physics recovery with limited data, which purely data-driven methods cannot achieve.

The major challenges, when working with unsteady flow-field data past moving bodies, arise in the temporal domain of the problem. An understanding of the dynamical behavior of the flow and its interaction with the moving body depends on long time integration of the flow-field, posing a challenge for PINNs as they do not train well on long-time domains~\cite{krishnapriyan2021characterizing}. Often, due to memory/storage constraints, the temporal resolution of the obtained data can be coarser than the minimum required Nyquist sampling criterion. In such a scenarios, standard interpolation techniques might not suffice for flow-field reconstruction at intermediate time stamps.
Moreover, it might not be feasible to re-run the simulations or experiments for the entire spatial domain when the region of interest is often a truncated spatial subdomain encompassing the near-field region and the moving body. Hence, a physics-informed continuous time-space interpolator like a PINN would be beneficial as it can work under data-sparse conditions~\cite{cai2021physics}, and with any arbitrary domain~\cite{raissi2020hidden}. Importantly, sparsity is a computational source of complexity for PINNs, whereas, flow-field aperiodicity necessitates long-time integration and thus strongly tied to the physics of the flow. Aperiodic/quasi-periodic flows past flapping foils are investigated by the community owing to their ability to enhance lift/thrust generation~\cite{khalid2018bifurcations, lewin2003modelling, platzer2008flapping, bose2018investigating, majumdar2022transition, shah2021investigating}.
Such flow-fields contain rich temporal scales although the flow-field might look seemingly regular. A rich spectral content might prove to be challenging for PINNs owing to their frequency bias issues. One way to mitigate this is to use Fourier embeddings~\cite{wang2021eigenvector}. However, it is impossible to pre-assess the frequency content of the flow-field, before experiments/simulations. Also, even with available data, it might be challenging to determine the spectrum if the temporal resolution is poor. 

Considering the above situations, effective training strategies need to be employed to handle the temporal domain complexity of different forms: (1) temporal sparsity, (2) long-time domain integration, and (3) rich spectral content in case of aperiodicity. 
A unified causal framework for soft/hard causality enforcement through weighting or time domain decomposition strategies was proposed recently~\cite{penwarden2023unified}. These sequential learning strategies, more than parallel training, can benefit from the transfer learnability of traditional neural networks. But this is applicable only when the networks train on equidistant time/spatial windows in the physical domain. 
Although it might seem promising to consider varied sizes of networks to train different temporal segments~\cite{jagtap2021extended}, transfer learning would only be beneficial when the time windows/segments chosen are the same, given that spatial features are often quite similar across temporal snapshots of periodic/aperiodic flows. 
This motivates the present study in which different sequential learning strategies have been explored, including time marching~\cite{krishnapriyan2021characterizing}, backward compatible training~\cite{mattey2022novel, malineni2025performance}, and sequential learning with transfer learning methods~\cite{tang2022transfer}. These have been investigated in the context of unsteady flows past a flapping airfoil. 

For flow past a plunging foil at a low Reynolds number having known kinematics and velocity flow-field data from IBM simulations (sparsely sampled in time), the MB-PINNs~\cite{sundar2024physics} have been extended here using sequential training strategies to obtain surrogate models. The aim of the surrogates is to perform velocity reconstruction and simultaneous pressure recovery. In light of the above discussed challenges, the scope, and the contributions of the present study are as follows.
\begin{itemize}
   \item To begin with, evaluation of the earlier proposed MB-PINN for velocity reconstruction and pressure recovery under different temporal domain complexities to identify its limitations. 
    \item Extension of MB-PINN to efficiently handle temporal domain complexities using time marching, and temporal decomposition based sequential strategies coupled with transfer learning.
    \item Detailed evaluation of the proposed methods under a fixed training budget over both periodic and quasi-periodic flow situations past a flapping body,   depicting the three temporal domain complexities discussed above. 
    \item Devising an effective preferential spatiotemporal sampling strategy to train sequential learning variants for accurate near-field pressure recovery. 
\end{itemize}
In the present investigation, the efficacy of the sequential training strategies, and computational efficiency have been  compared under a fixed training budget.
A physics-based sampling,has been used to train the networks in a data-efficient manner. 
The general structure of the paper is as follows. In Section~\ref{sec:seqIBAPINN}, the problem setup, immersed boundary-aware framework, and sequential learning framework are discussed in the context of pressure recovery from velocity data. 
The IBM and ALE-based CFD solver settings are used for data generation, and details of the training and testing data sets are given in Section~\ref{sec:database}. Results and discussion for the periodic and aperiodic test cases are presented in sections~\ref{sec:periodicresults}, and \ref{sec:quasiperiodicresults}, respectively. Finally, the key conclusions and further works are outlined in Section~\ref{sec:conclusion}.

\section{Sequential PINNs for unsteady flows past moving bodies}\label{sec:seqIBAPINN}
The details of the plunging airfoil system, modeling of the unsteady viscous flow-field, formulation of sequential learning methods with MB-PINN as the underlying backbone, and training methodology are presented in the following subsections.

    \begin{figure}[!t]
    \centering
    \includegraphics[width = 0.8\linewidth]{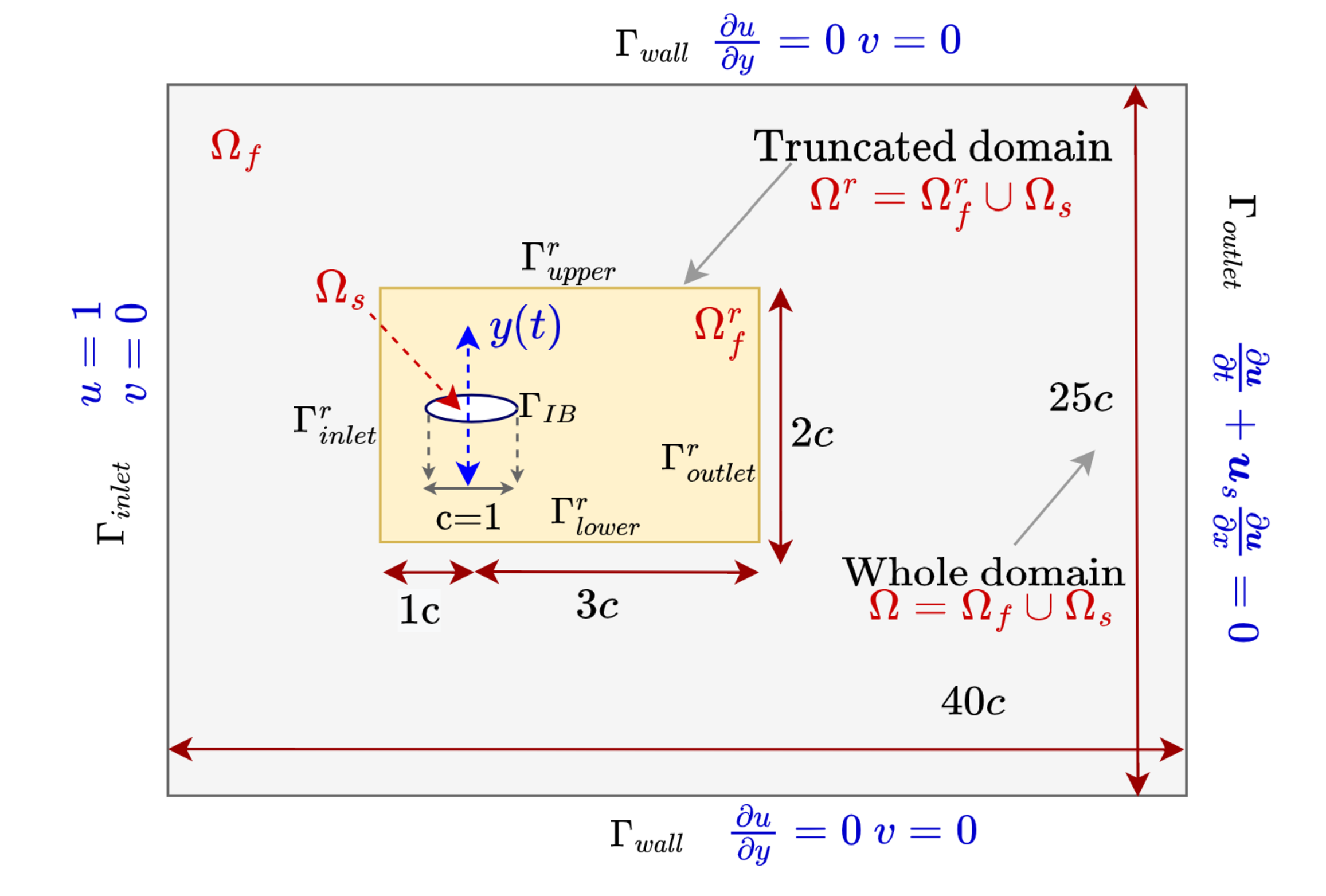}
    \caption{Schematics of the problem setup: computational domain $\Omega$ is chosen for the IBM solver, and the truncated domain $\Omega^r$ is chosen for the surrogate model. $\Omega_f$ and $\Omega_f^r$ are the fluid regions excluding the solid boundary $\Gamma_{IB}$ and solid region $\Omega_s$ at any given time instant. $\Gamma_{inlet}^r,$ $\Gamma_{outlet}^r,$ $\Gamma_{upper}^r$ and $\Gamma_{lower}^r$ are the inlet, outlet, upper and lower boundaries of the truncated domain $\Omega_r$, respectively.}
    \label{fig:problemsetup}
    \end{figure}

 \subsection{The flapping foil system}
	In the present study, unsteady incompressible flow past a plunging rigid elliptic airfoil is considered as a canonical problem to evaluate the efficacy of the proposed methods. This could be considered as a good test for moving body flows owing to the associated complex flow-features involving strong leading and trailing edge vortices, their interactions as well as the potentially rich dynamics. 
    
    The flow is assumed to be two-dimensional and governed by a low Reynolds number.  
	The plunging foil of unit chord length ($c = 1$) is considered to follow an imposed harmonic kinematic motion. It is subjected to uniform free-stream; see figure~\ref{fig:problemsetup}(a) for a schematic of the problem. 
	The kinematic model is as follows
	\begin{align} 
	{y}({t}) &= h_a \cos(2\pi f_h {t} + \phi), \;\; \mbox{and} \label{eq:kin1}
	\\\dot{{y}}({t}) &= -2\pi f_h h_a \sin(2\pi f_h {t} + \phi). \label{eq:kin2}
	\end{align}
	Here, $t$ is dimensional time; $h_a$ and $f_h$ are the plunging amplitude and frequency, respectively. Here, $\phi$ is the kinematic phase parameter that determines the initial body position at $t = 0$.  Aligning with earlier literature~\cite{lewin2003modelling,khalid2018bifurcations}, further discussions will be in the context of non-dimensional amplitude $h = h_a/c$ and reduced frequency $k = 2\pi f_h c / U_{\infty},$ where, $U_{\infty}$ is the free stream velocity.

	\subsection{Modeling of the Unsteady flow}\label{sec:IBM}
	
	The flow around the flapping foil is governed by the incompressible Navier-Stokes (N-S) equations given in its non-dimensional form by
	\begin{align}
	\frac{\partial \boldsymbol{u}}{\partial t} + \nabla.({\boldsymbol{u}}{\boldsymbol{u}}) &= -{\nabla}{p} + \frac{1}{Re} {\nabla}^2{\boldsymbol{u}}, \label{eq:NSEq1} \\
	{\nabla}.{\boldsymbol{u}}&= 0. \label{eq:NSEq2}
	\end{align}
	Here, ${\boldsymbol{u}}$ represents the non-dimensional velocity vector in the $x$-$y$ space with $u$ and $v$ being the $x$ and $y$ components of $\boldsymbol{{u}},$ respectively, and ${p}$ is the non-dimensional pressure;  $\displaystyle Re = \frac{U_{\infty}c}{\nu}$ indicates the Reynolds number with $\nu$ being the kinematic viscosity.
	
\begin{figure}[!t]
    \centering
    \includegraphics[width=\linewidth]{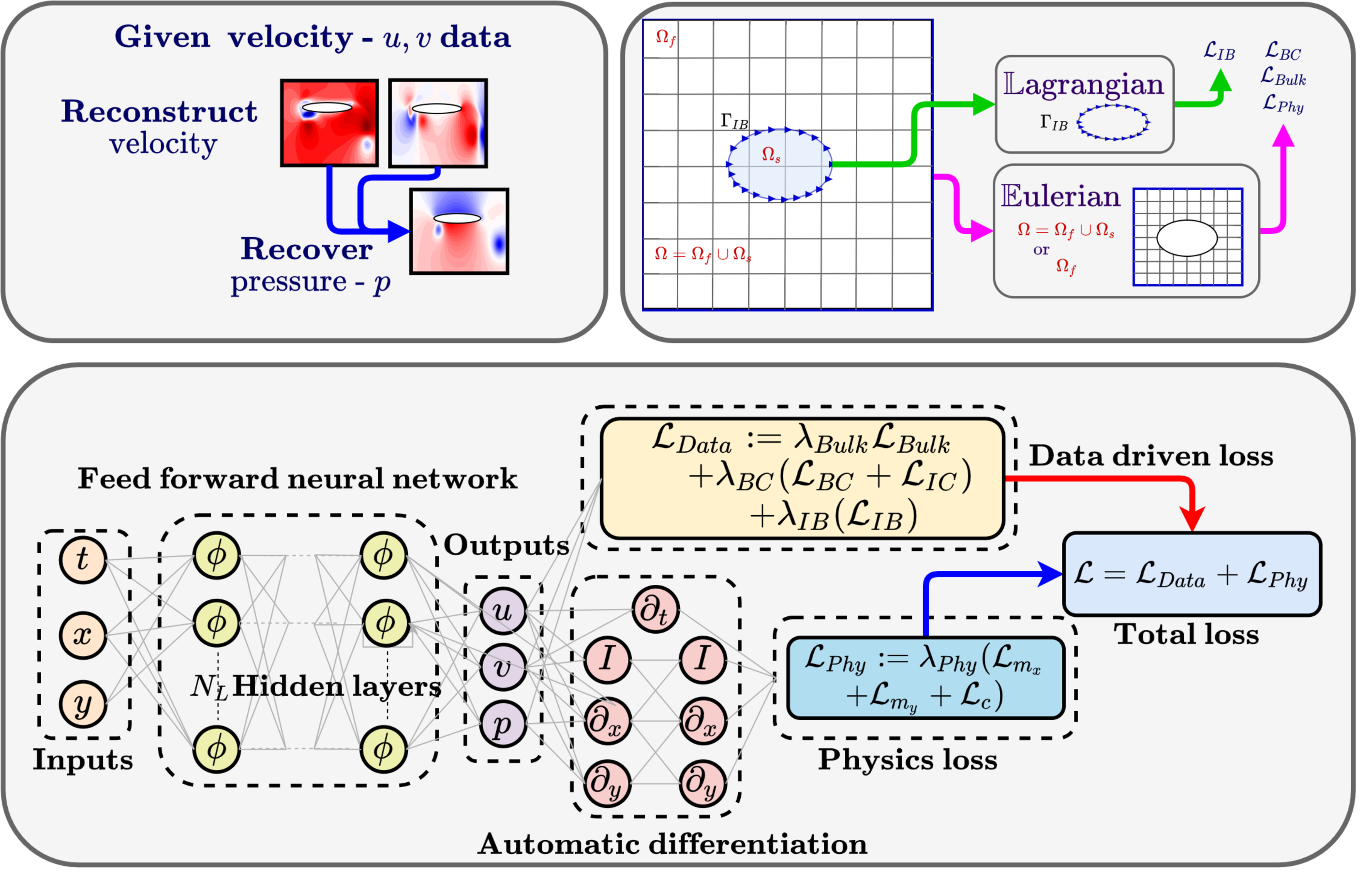}
    \caption{A schematic depicting the MB-PINN formulation under the immersed boundary aware (IBA) framework for pressure recovery from velocity data. }
    \label{fig:ibaframework}
\end{figure}

  In the IBM approach, the immersed solid boundary $\Gamma_{IB}$ is represented by a set of Lagrangian markers (figure~\ref{fig:ibaframework}) which do not conform exactly with the Eulerian fluid grid. Importantly, at the solid boundary $\Gamma_{IB}$, a no-slip boundary condition is to be best satisfied such that $\boldsymbol{u}(\boldsymbol{x} = \Gamma_{IB},t) = \dot{y}(t)$. Since this cannot be directly enforced on $\Gamma_{IB}$ in the IBM, a momentum forcing term $\boldsymbol{f}$ with $f_x$ and $f_y$ as the respective $x$ and $y$ components must be added in equation~(\ref{eq:NSEq1}). $\boldsymbol{f}$ is evaluated such that the no-slip boundary condition is satisfied on $\Gamma_{IB}$ using appropriate mapping and interpolation between the Eulerian and Lagrangian points. Additionally, a source/sink term $q$ is added in equation~(\ref{eq:NSEq2}) to strictly satisfy the continuity equation. Note that both $\boldsymbol{f}$ and $q$ are non-zero inside $\Omega_s$ bounded by $\Gamma_{IB}$. 
	Thus the governing equations  take the form below, 
	\begin{align}
	\frac{\partial \boldsymbol{u}}{\partial t} + \nabla.({\boldsymbol{u}}{\boldsymbol{u}}) &= -{\nabla}{p} + \frac{1}{Re} {\nabla}^2{\boldsymbol{u}} + \boldsymbol{f}, \label{eq:IBMEq1} \\
	{\nabla}.{\boldsymbol{u}} - q &= 0. \label{eq:IBMEq2}
	\end{align}
	For more details on the IBM solver, please refer to~\cite{majumdar2020capturing,shahperformance}. An in-house GPU parallelized unsteady flow solver~\cite{majumdar2020capturing} has been employed here to solve equations~(\ref{eq:IBMEq1}) and~(\ref{eq:IBMEq2}).

	Pressure field data was not stored under our current IBM framework~\cite{majumdar2020capturing}, as pressure data was not needed for computing the aerodynamic loads. 
 For a rigorous testing of the pressure recovery capability of the PINN models, a well-validated ALE-based solver was used here to generate and test the pressure data, as was also done in our study~\cite{sundar2024physics}.  
 Note that in the ALE technique, the governing Navier-Stokes equations are solved without the additional forcing or mass source/sink terms. ALE-based simulations were performed using OpenFOAM foam-Extend's icoDyMFoam solver~\cite{jasak2007openfoam,majumdar2020capturing}. For aperiodic flows minor differences in field conditions could potentially grow, resulting in 
 ALE and IBM obtained solutions slightly differing from each other mainly owing to differences in the underlying numerical methods~\cite{majumdar2020capturing}. This could potentially bring small discrepancies in the aerodynamic loads in the aperiodic case. Thus, a straightforward cross-comparison of model predictions might be difficult for the aperiodic case.   

 \begin{figure}[!t]
    \centering
    \includegraphics[width=\linewidth]{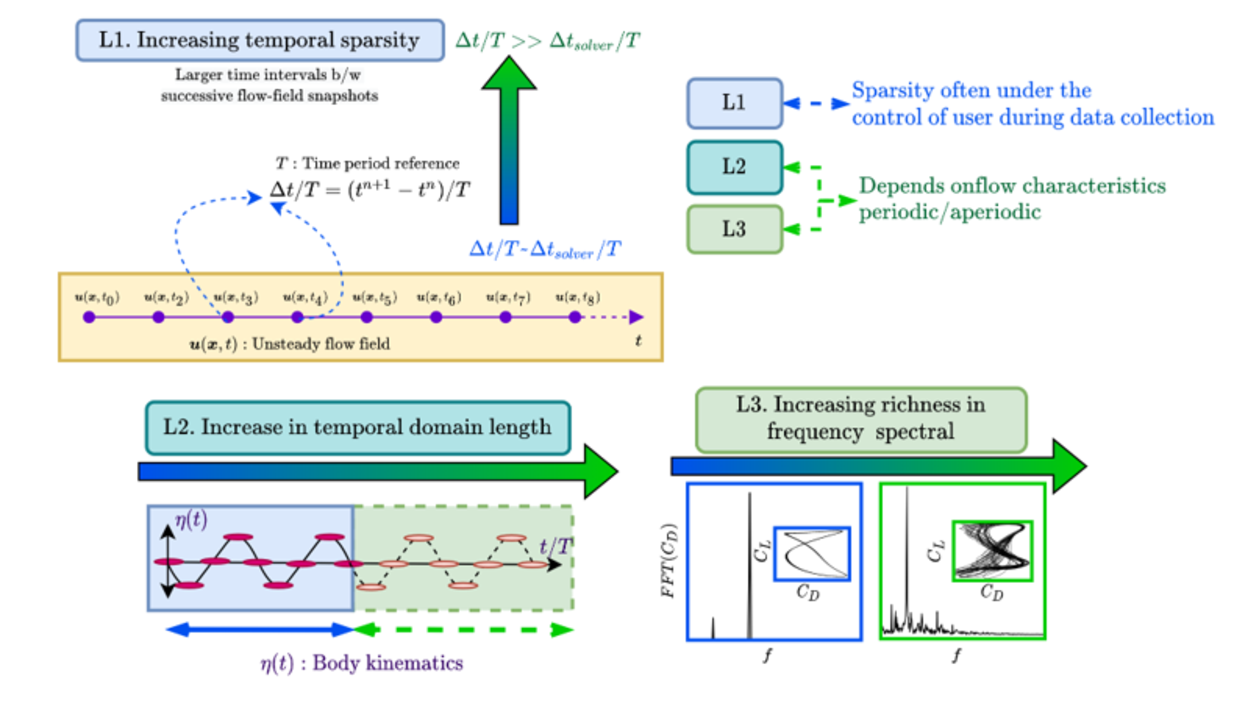}
    \caption{Schematic depicting different computational and physical sources of temporal domain complexities with  application to unsteady flow past a sinusoidally plunging airfoil.}
    \label{fig:temporalcomplexities}
\end{figure}

\subsection{Immersed boundary aware framework} 
Surrogate modeling of unsteady flows past moving bodies using a fixed frame of reference is beneficial as it can allow handling multiple moving bodies, and possible deforming bodies. Recently,  Sundar {\it et al.}~\cite{sundar2024physics} proposed an  immersed boundary  aware (IBA) framework of PINNs which considered a fixed Eulerian frame of reference and was inspired by the benefits of the immersed boundary method. 
Note that, it would not be feasible to enable coordinate transformations when multiple moving bodies or flexible bodies are involved. The reported IBA framework in ~\cite{sundar2024physics} alleviates the need for transformations of the computational domain to body-attached frame of reference. The fluid velocity data points were obtained on a fixed Eulerian grid, whereas, the solid body was immersed in the Eulerian grid and described using a set of Lagrangian markers. Pressure recovery from velocity data in the modeling of unsteady flows past moving boundaries (see figure~\ref{fig:ibaframework}) was one of the major outcomes. Sundar {\it et al.}~\cite{sundar2024physics} showed that although the data was generated using an IBM solver, in a scenario where body position and velocity are known {\it a priori}, the NS-equations-based formulation, MB-PINN, was best suited for pressure recovery. In this scenario, it was also shown that, the IBM-based formulation MB-IBM-PINN could perform at par with MB-PINN only when the solid region was discarded from physics loss computation and that too with an additional computational overhead. However,  when body position, shape, and velocity were not exactly known and the solid region could not be determined {\it a priori},   MB-IBM-PINN could be useful. In the NS based MB-PINN formulation, $\boldsymbol{f}$ was strictly zero outside $\Gamma_{IB}$ everywhere in $\Omega_f$; $q$ was  non-zero just outside $\Gamma_{IB}$ at grid cells that were partially cut by $\Gamma_{IB}$ and zero everywhere else in $\Omega_f$~\cite{sundar2024physics}.  

\noindent 
The detailed loss formulation for the standard MB-PINN  is as follows~\cite{sundar2024physics} 
	\begin{align}
	\mathcal{L}_{Data}:&= \lambda_{Bulk}\mathcal{L}_{Bulk} + \lambda_{Inlet} (\mathcal{L}_{Inlet}) + \lambda_{IB}\mathcal{L}_{IB}, \;\;\mbox{and}\label{eq:dataloss-MBPINN}
	\\\mathcal{L}_{Phy}:&= \lambda_{fluid}(\mathcal{L}_{{m_x}} + \mathcal{L}_{{m_y}} + \mathcal{L}_{c}).\label{eq:phylossnosplit-MBPINN}
	\end{align}
	The loss components,  $\mathcal{L}_{Bulk}\;\mbox{and} \;\;\mathcal{L}_{Inlet}$ correspond to predicted interior bulk velocity data and inlet velocity boundary conditions on the Eulerian grid. The additional no-slip velocity boundary condition loss, $\mathcal{L}_{IB}$, is computed directly on the solid boundary,  $\Gamma_{IB}$, described by a set of Lagrangian markers. The coefficients, $\lambda_{*}$, for $* \in \{\text{Bulk, Inlet, IB}\}$ are the weighting coefficients of the bulk data loss, inlet, and no-slip boundary condition loss components, respectively. Here, $\lambda_{fluid}$ is the weight of the physics loss component which operates only in the fluid region by design.  

	A general mathematical formulation of the above-mentioned data-driven and physics loss components is as follows
	\begin{gather}
	\mathcal{L}_{*}:= \frac{1}{N_{*}}\sum_{i = 1}^{N_{*}}\|\boldsymbol{u}(\boldsymbol{x}_{*}^i,t^i) - \hat{\boldsymbol{u}}(\boldsymbol{x}_{*}^i,t^i) \|^2_{L_2}
 \\\mathcal{L}_{{@}} := \frac{1}{N_{Phy}}\sum_{i = 1}^{N_{Phy}}\|r_@^x(\boldsymbol{x}_{Phy}^i,t^i)\|^2_{L_2}
	\end{gather}
	where, ${\boldsymbol{\hat{u}}}$ is the true velocity data generated for training by the CFD simulation, whereas, ${\boldsymbol{u}}$ is the network-predicted velocity flow-field data. 
 The spatial and temporal points are represented by $(\boldsymbol{x}, t)$ with $\boldsymbol{x} \in \Omega^r$ and $t \in [0,T]$, respectively. The subscripts shown in the loss expressions for the spatiotemporal points are to indicate independence in sampling the select points of the particular loss component. Here, $(\boldsymbol{x}_{*}^i,t^i)$ for $i = \{1,\cdots N_{*}\}$ corresponds to the set of velocity data points in the respective fluid interior bulk region, at the inlet or on the solid boundary,  respectively. 
 The physics-informed loss components,  $\mathcal{L}_{{m_x}},\;\mathcal{L}_{{m_y}}$, and  $\mathcal{L}_{c}$ correspond to the mean squared errors of $x$ and $y$ momentum equations (Eqn.~\ref{eq:NSEq1}) and continuity equation (Eqn.~\ref{eq:NSEq2}) residuals $\boldsymbol{r_@}$, respectively. Here, the subscript $@$ denotes the different physics loss components.
 
The following forms of residuals are used for MB-PINN
\begin{gather}\label{eq:residuals}
r_m^x(\boldsymbol{x}_{Phy}^i,t^i) = u_t + uu_x + vu_y - p_x - 1/Re(u_{xx} + u_{yy}),	
\\r_m^y(\boldsymbol{x}_{Phy}^i,t^i) = v_t + uv_x + vv_y - p_y - 1/Re(v_{xx} + v_{yy}),\;\;\mbox{and,}
\\r_c(\boldsymbol{x}_{Phy}^i,t^i)= u_x + v_y. 
\end{gather}
Here, $(\boldsymbol{x}_{Phy}^i,t^i)$ for $i = \{1,\cdots N_{Phy}\}$ is the set of points from the interior of the domain $\Omega^r$, where the governing equation residuals are evaluated to compute the physics loss, $\mathcal{L}_{Phy}.$

As already stated, the focus of the present work is on  different temporal domain complexities and recovery of pressure from velocity data under them. Though the fluid velocity data are not well resolved in time, it is 
 assumed that the body position and velocity are available in a time-resolved manner. Throughout the study, sequential learning based extensions of the MB-PINN will be considered for the pressure recovery problem. The details of the framework and a general loss formulation are discussed in the next section. 

\subsection{Network architecture and loss formulation}
As mentioned in the introduction, physical and numerical complexities related to the temporal dynamics of the system (see figure~\ref{fig:temporalcomplexities}), such as, (a) temporal sparsity, (b) long-time integration, and (b) rich temporal spectrum, need to be handled while training PINNs for moving body flows in question. In the absence of any prior knowledge of the frequency content of the flow, the negative impact of these temporal complexities can be alleviated by turning to sequential learning. 
Extending MB-PINN~\cite{sundar2024physics} inspired by the unified scalable causal framework~\cite{penwarden2023unified}, a general loss formulation is proposed, which under suitable conditions morphs to sequential learning variants of the MB-PINN model. 

For spatiotemporal coordinates, $\{(\boldsymbol{x}, t)\}_{i = 1}^N \in \Omega_{\boldsymbol{x}} \times \Omega_t, $ $N_d$ domain decompositions can be obtained such that,  $\Omega_{\boldsymbol{x}}\times \Omega_t = \bigcup_{i = 1}^{N_d}\Omega_{\boldsymbol{x}}^i\times\Omega_t^i$, with common stationary or moving interfaces $\{\partial\Omega_{ij}\}_{i = 1, j=1}^{N_d, N_d}  \forall i\neq j.$ 
The temporal domain is broken down into subdomains and these interfaces/common data points are only considered in the temporal domain context. 
 Now, an array of $N_{nets}$ number of neural network backbones, $\{_{\#_j}\mathcal{N}(\theta_j
)\}_{j = 1}^{N_{nets}}$, can be defined on each subdomain or a combination of subdomains, such that $\#_j$ is a feed-forward neural network. One can choose variants of a feed-forward neural network as well,  such as the modified MLP (mMLP)~\cite{wang2020understanding}, or, a Fourier feature-based multiscale network~\cite{wang2021eigenvector}. A simple feed-forward network is considered in the present study. Within each subdomain, multiple sets of corresponding spatio-temporal coordinates and target variables are used for training which have been described in the following. 

For any $i^{\text{th}}$ subnetwork being trained, the overall loss formulation is mathematically expressed as:
 \begin{align}
     \mathcal{L}(\theta_i):&= \mathcal{L}_{Data}(\theta_i) + \mathcal{L}_{Phy}(\theta_i) + \mathcal{L}_{Backward}(\theta_i),
    \\\mathcal{L}_{Data}(\theta_i):&= \lambda_{Bulk}^i\mathcal{L}_{Bulk}(\theta_i) + \lambda_{inlet}^i\mathcal{L}_{inlet}(\theta_i) + \lambda_{IB}^i\mathcal{L}_{IB}(\theta_i)
	\\\mathcal{L}_{Phy}(\theta_i):&= \lambda_{Mom}^i(\mathcal{L}_{{m_x}}(\theta_i) + \mathcal{L}_{{m_y}}(\theta_i)) + \lambda_{Con}^i(\mathcal{L}_{c}(\theta_i)),
  \\\mathcal{L}_{Backward}(\theta_i) &= \lambda^i_{Back} \sum_{j = 1}^{N_d^{Data} - 1}\mathcal{L}_{Bulk}^{j}.
 \end{align}

Here, $N_d^{Data}, N_d^{Phy}$ are the number of subdomains with data and physics losses, whereas $N_{\mathcal{I}}$  corresponds to the number of common interfaces of the subdomains and $N_\leftarrow^{Data^-}$ corresponds to the number of networks with backward compatible losses. The loss coefficients, $\lambda_{*}^i$, indicate that one can choose different weights for each $i^{th}$ subdomain independently. These weights can be either determined globally or locally (i.e. in a point-wise manner)~\cite{mcclenny2023self, xiang2022self, wang2020understanding, anagnostopoulos2023residual}. In the present study, a global physics loss relaxation is adopted aligning with the earlier works~\cite{lucor2022simple, sundar2024physics}.

\begin{figure}[!t]
    \centering
    \includegraphics[width=\linewidth]{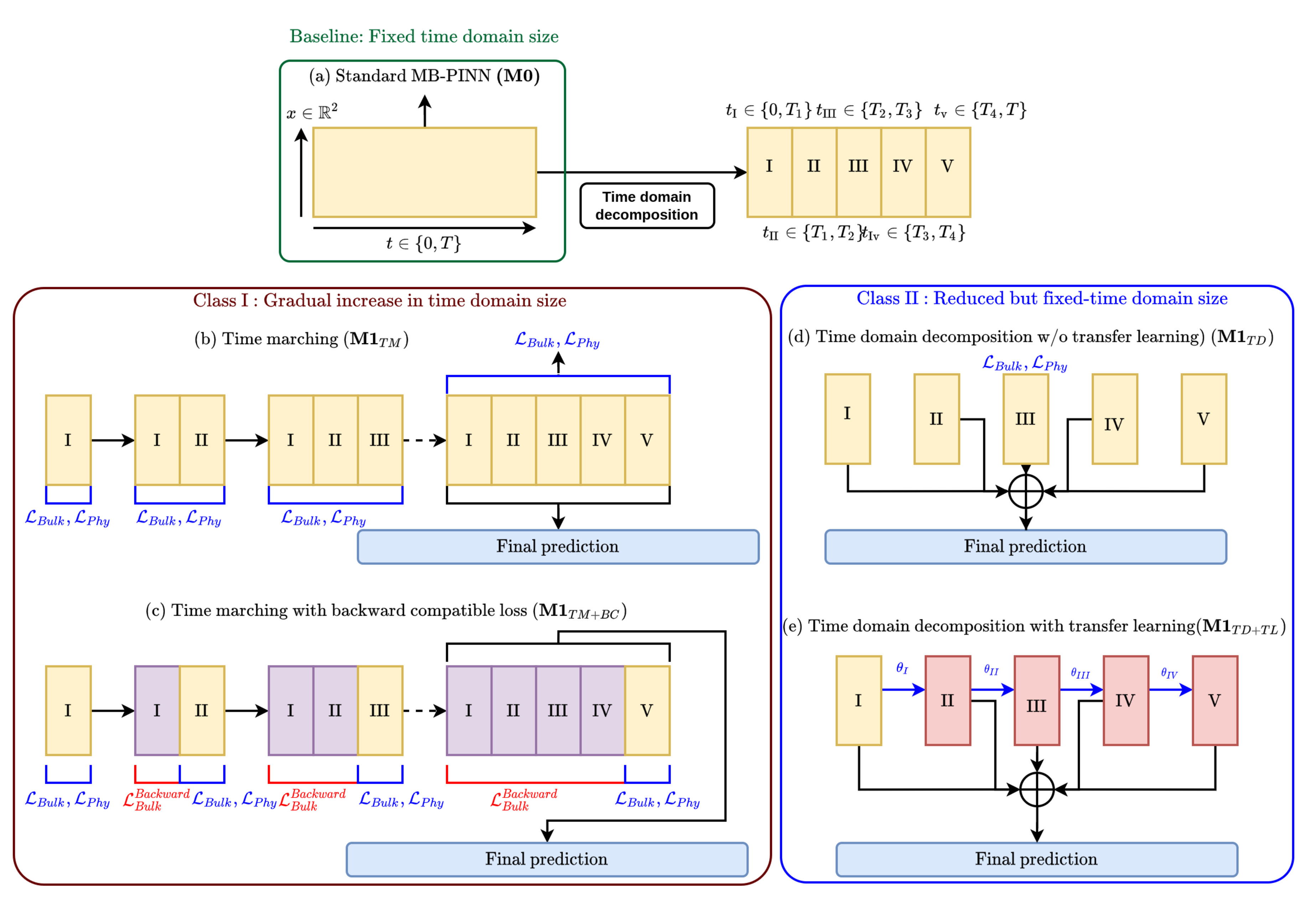}
    \caption{Schematics  of  different types of sequential learning variant of MB-PINN considered. The entire time domain is partitioned into  subdomains (five, as an example) of equal length. The green, red, and blue outlined boxes refer to broad classification in terms of the time domain size the model sees during training;  ClassI: $\mathbf{M1}_{TM}$ and $\mathbf{M1}_{TM+BC}$ fall under this  for which a single network sees a gradual increase in the time domain by appending the subdomains one by one during training; Class II:  $\mathbf{M2}_{TD}$ and  $\mathbf{M2}_{TD+TL}$ fall under this class for which multiple networks form the model, each network inheriting the data of a subdomain.}
    \label{fig:seqlearningframework}
\end{figure}

Different MB-PINN variants are obtained from the generalized loss formulation shown above (see figure~\ref{fig:seqlearningframework} for a detailed schematic and table~\ref{tab:seqmethods}). In addition to the baseline standard MB-PINN shown in the figure~\ref{fig:seqlearningframework}, the sequential learning variants of MB-PINN have been  further classified into two broad categories: Class I, and Class II,  depending on whether the time domain is gradually increased in size, or decomposed into small subdomains during training (figure~\ref{fig:seqlearningframework}).

\paragraph{\em{\textbf{Baseline} ($\mathbf{M0}$)}
}  
In this case, MB-PINN is trained over the entire  time domain  effectively as a single domain and a single network under consideration.  As a result, $N_{d}^{Data} = N_{d}^{Phy} = 1$ (see figure~\ref{fig:seqlearningframework}(a)). 

\subsubsection{Class I Models}
In this variant, starting with the first subdomain, a {\it single} network ($N_{nets} = 1$) is trained in stages, and with each passing training stage the time domain size increases gradually by the addition of the subsequent temporal subdomain. Once trained, only this {\it single} network is necessary to infer the predictions. 

\paragraph{\em{\textbf{Time marching }($\mathbf{M1}_{TM}$)}} In this approach,  the time domain is decomposed into $N_d^{data} = N_{d}^{Phy}$ sequentially ordered input time slabs/subdomains for simplicity.  Note that while the model size is fixed, the time domain size is gradually increased (see figure~\ref{fig:seqlearningframework}(b)).

\paragraph{\em{\textbf{Time marching + Backward compatible training}($\mathbf{M1}_{TM + BC}$)}} This approach incorporates elements from the work of Mattey {\it et al.} ~\cite{mattey2022novel}.  $\mathbf{M1}_{TM+BC}$ is trained such that $\mathcal{L}_{Data}(\theta_i)$ and $\mathcal{L}_{Phy}(\theta_i)$ 
 are computed over $i = 1,2 \cdots N_d^{data} (= N_d^{Phy})$ temporal subdomains. In addition, backward compatible data losses ($\mathcal{L}_{Backward}$) are introduced to fit the predictions obtained over the previous time slabs using the network used in the current time slab. (see figure~\ref{fig:seqlearningframework}(c)). 
\subsubsection{Class II Models}
In this category, the size of the time domain is reduced by splitting the time domain into smaller subdomains and training individual networks over each. Post-training,  these networks need to be strung together to obtain the predictions over the entire time domain. 
While this increases the number of networks to be trained, due to reduced time domain size for each network potentially  a smaller network can be trained with a lower computational time. 

\paragraph{\em{\textbf{Time domain decomposition }(($\mathbf{M2}_{TD}$) + \textbf{Transfer learning} ($\mathbf{M2}_{TD+TL}$))}}
The time domain is decomposed into $N_d^{data} = N_{d}^{Phy}$ subdomains in this approach as well. But, unlike $\mathbf{M1}_{TM}$ or $\mathbf{M1}_{TM+BC}$, there are $N_{nets} = N_d^{data}$ number of networks trained subsequently over each subdomain (see figures~\ref{fig:seqlearningframework}(d) and (e)). Note that  $\mathbf{M2}_{TD}$ does not incorporate transfer learning of weights, while $\mathbf{M2}_{TD + TL}$ does so. The final weights of the trained network from the previous subdomain are used (transferred) as initialization for training over the subsequent subdomain consecutively, hence termed as a transfer learning approach. 
\begin{table}[!htbp]
    \centering
    \caption{Overview of the standard and different sequential learning variants of MB-PINN }
    \label{tab:seqmethods}
    \resizebox{\textwidth}{!}{\begin{tabular}{c|c|c|c|c}
        \hline
         Model & Time Marching & Backward compatible & Time domain decomposition & Transfer learning \\
         \hline
         $\mathbf{M0}$ &  $\tikzxmark$ & $\tikzxmark$ & $\tikzxmark$ & $\tikzxmark$ \\
         $\mathbf{\mathbf{M1}}_{TM}$ &  $\tikzcmark$ & $\tikzxmark$ & $\tikzxmark$ & $\tikzxmark$ \\
         $\mathbf{\mathbf{M1}}_{TM+BC}$ &  $\tikzcmark$ & $\tikzcmark$ & $\tikzxmark$ & $\tikzxmark$ \\
         $\mathbf{M2}_{TD}$ & $\tikzxmark$ & $\tikzxmark$ & $\tikzcmark$ & $\tikzxmark$ \\
         $\mathbf{M2}_{TD+TL}$ &  $\tikzxmark$ & $\tikzxmark$ & $\tikzcmark$ & $\tikzcmark$ \\
         \hline 
    \end{tabular}}
\end{table}

The loss function is optimized over the network parameters, $\theta = \{\theta_i\}_{i = 1}^{N_{nets}}$, using an ADAM optimiser~\cite{kingma2014adam}.  
For a fair comparison, baseline, Class I and II models (sub-networks) are trained under a fixed training budget of 1.5e06 iterations with a step decay of learning rate after every 5e05 iterations, starting from $\eta = 1e-03$, $5e-04$ over the second step, and finally to $1e-04$.

\section{Database generation}\label{sec:database}
The test cases to evaluate the  proposed methods under temporal domain complexities are outlined here. The details of the training and testing databases are then presented. 
\subsection{Test cases}
\begin{figure}[!b]
    \centering
    \includegraphics[width=\linewidth]{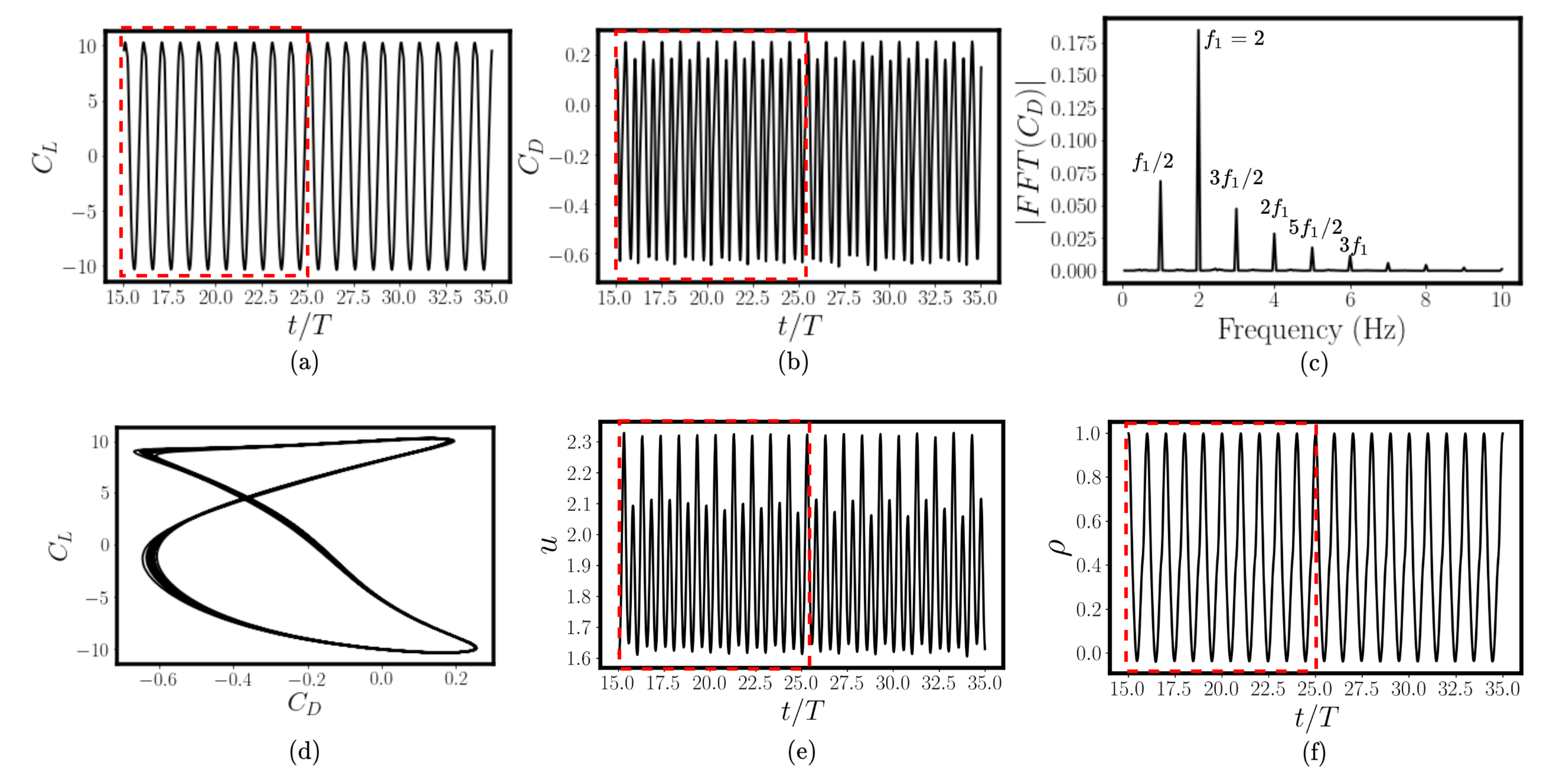}
    \caption{Dynamics and data analysis of the periodic case: (a-b) coefficients of lift ($C_L$) and drag ($C_D$) with  time, (c) $C_D$ Fourier spectrum, (d) $C_L-C_D$ phase portrait, (e) streamwise velocity ($u$) at probe location $x/c = 2.5,$  and,  (f) corresponding streamwise velocity correlation ($\rho$) as a function of time ($t/T$).}
    \label{fig:periodic_flow_characterisation}
\end{figure}
\begin{figure}[!b]
    \centering
    \includegraphics[width=\linewidth]{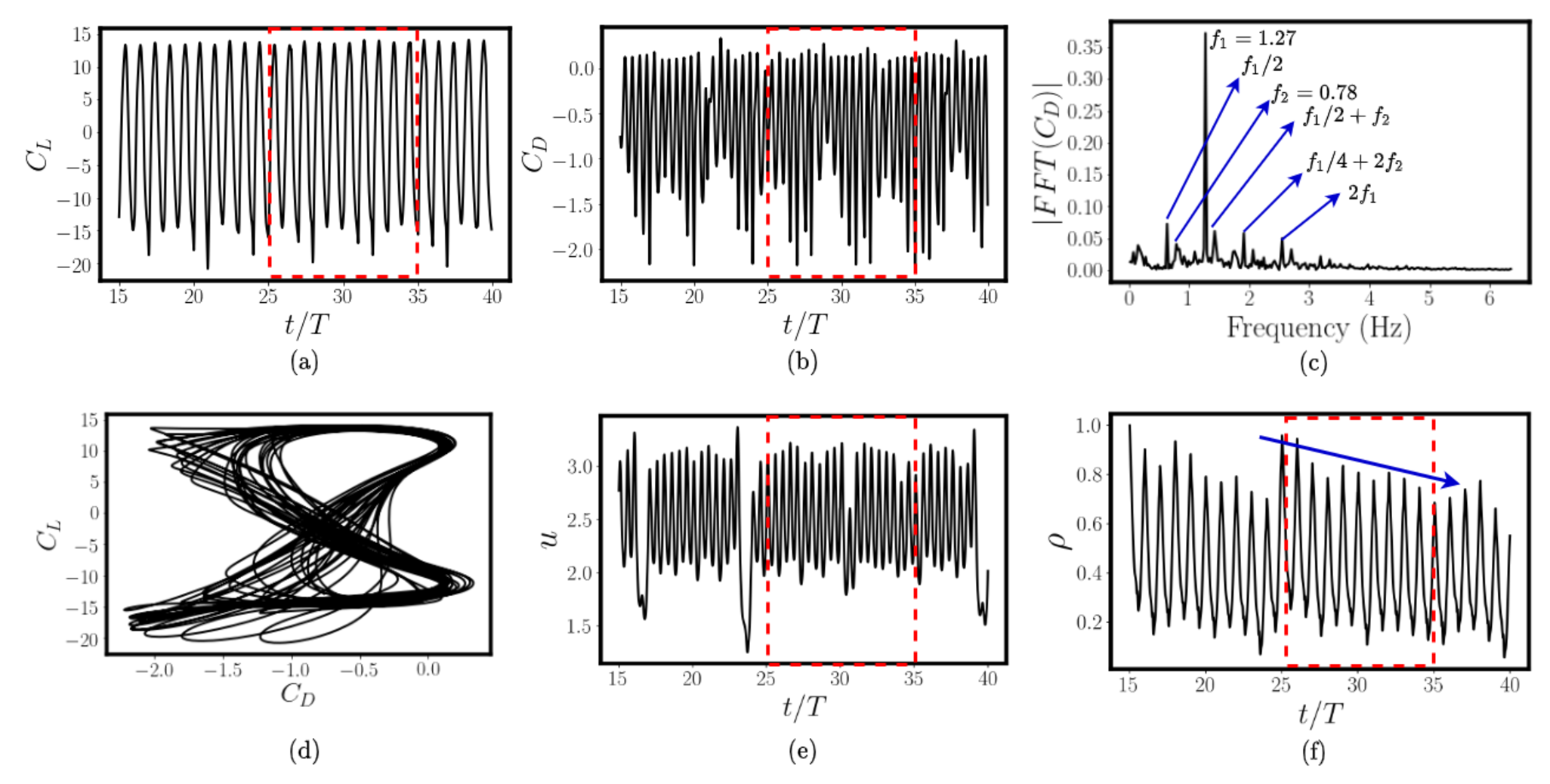}
    \caption{Dynamics and data analysis of the quasi-periodic case: (a-b) lift and drag coefficient time histories, (c) $C_D$ Fourier spectrum, (d) $C_L-C_D$ phase portrait, (e) streamwise velocity ($u$) at probe location, $x/c = 2.5,$ and, (f) corresponding streamwise velocity correlation ($\rho$) as a function of time ($t/T$).}
    \label{fig:quasiperiodic_flow_characterisation}
\end{figure}

Two unsteady flow scenarios were considered in the present study  to evaluate the performance of the sequential learning variants  under different temporal domain complexities mentioned earlier. 
In the first scenario, the flow was simulated at $Re = 500,$ reduced frequency, $k = 2\pi$ and plunge amplitude, $h = 0.16,$ corresponding to $kh = 1.0$~\cite{khalid2015analysis}. The flow-field in this case was completely periodic. The second scenario involved a more challenging case of a quasi-periodic flow as in~\cite{majumdar2020capturing}. It was simulated with a slightly lower Reynolds number $Re = 300$ at $k = 4$,  $h = 0.4125$ resulting in  $kh = 1.65$. The aperiodic case is more challenging because of two main reasons: lack of repeatability of the flow-field, and its spectral richness. Note that, the combination of a relatively large plunging amplitude and lower Reynolds number make the vortices larger  leading to a relatively larger domain extent than in the periodic case if one were to capture at least two vortex couples. All the unsteady flow simulations were carried out using an in-house IBM solver~\cite{majumdar2020capturing,sundar2024physics}.  IBM based solvers in general are ideal for handling large movements of solid boundaries as they do not require mesh movement due to which the accuracy could suffer as in some other mesh conformal approaches, like the ALE~\cite{sarrate2001arbitrary}. Simulation data were coarsened in space-time for training (more on this will be discussed in Section 3.3), and the original high-resolution data were used for testing. To establish the dynamical states of the associated flow fields, a detailed characterization in terms of the aerodynamic loads and the stream-wise correlation of the $x-$ component of velocity has been shown in figures~\ref{fig:periodic_flow_characterisation} and~\ref{fig:quasiperiodic_flow_characterisation}. The time regular periodic case was chosen as a good test case for a thorough evaluation of the proposed variants under temporal sparsity, as well as the challenge of training over large time domains. Based on these evaluations, the most suited variants was identified and was then evaluated against the more challenging quasi-periodic case.  

For the quasi-periodic situation,  the frequency spectrum of the load (see in figure~\ref{fig:quasiperiodic_flow_characterisation}(c)) exhibits incommensurate frequencies. The presence of the incommensurate second frequency is not related to the input kinematics and is a direct result of the non-linearity in the flow governing equations. Moreover, the streamwise correlation of axial velocity component $u$ at a probe location $x/c = 2.5c$ (see figure~\ref{fig:quasiperiodic_flow_characterisation}(c)) shows a decreasing trend with an oscillatory behavior indicating quasi-periodicity~\cite{majumdar2020capturing}. Both case studies are expected to serve as benchmarks to evaluate the performance of standard (baseline) and sequential MB-PINN variants.

 \subsection{Vorticity cutoff based sampling}
 In order to improve the data efficiency of the MB-PINN model, a vorticity cut-off-based spatial undersampling strategy (VS) has been used in the present study as was proposed in~\cite{sundar2024physics}. Here, with the given bulk velocity data $\{\boldsymbol{u}(\boldsymbol{x}^i, t^i)\}_{i = 1}^{N_{Bulk}},$ vorticity values are  calculated as,  $\omega  = \nabla \times \boldsymbol{u}$.  Since vorticity can be considered a proxy for regions with strong flow-field gradients, a suitable cutoff $|\omega| = |\omega^*| > 0$, needs to be chosen to differentiate between the strong and weak gradient regions. Any desired percentage ($S_{|\omega |< |\omega^*|}$ and $S_{|\omega |\geq |\omega^*|}$) of bulk data points can then be sampled from these regions respectively. 
 
\begin{figure}
    \centering
    \includegraphics[width=\linewidth]{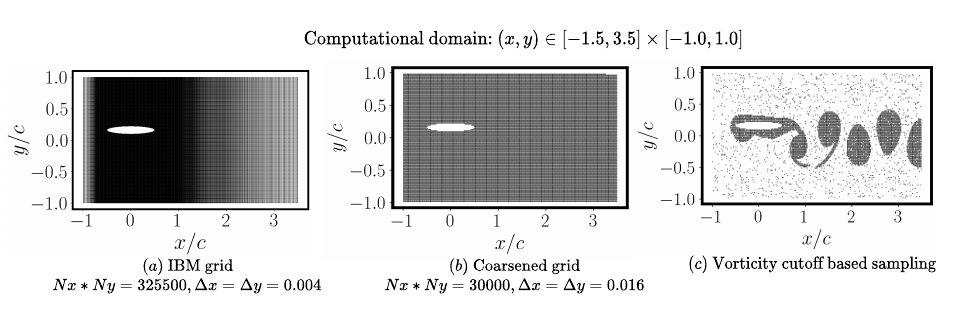}
    \caption{Computational domain and grids used for  training and testing for the periodic case; (a)  truncated high-resolution grid for IBM on which the models are tested, (b) ~4x coarsened grid with ~10x lower number of data points than in the high-resolution case, (c ) vorticity cut off based undersampled grid.}
    \label{fig:pgrids}
\end{figure}

\subsection{Details of training and testing datasets}

\begin{table}[!t]
	\centering
	\caption{ Details of the training and testing database for the domain considered for the periodic case. CI-*: coarse interpolated dataset; Ref-IBM and Ref-ALE:  high-resolution testing data sets generated using IBM and ALE solver, respectively. Further details on `Ref-*' datasets can be found in ~\cite{sundar2024physics}.}
	\begin{tabular}{ccccccc}
		\hline
		\textbf{Datasets} & $N_x$ & $N_y$ & $N_t^{Bulk}$& $(\Delta t/T)_{Phy}$ &$N_{Bulk}$ & $N_{Phy}$  \\
		\hline
            \multicolumn{7}{c}{Training datasets}\\
            \hline
            CI-11 & 270 & 120 & 11 & \multirow{2}{*}{0.025}& 3.564e05  & \multirow{2}{*}{2.5805e06}\\
            CI-S5-11 & - & - & 11 &  & \textbf{7.104e04}&\\
            \hline
            \multicolumn{7}{c}{Testing datasets}\\
            \hline
            Ref-IBM & 651 & 500 & \multirow{2}{*}{81} & - & 2.6365e07 &-\\
		Ref-ALE & - & - &  & - & 4.05e06 &-\\
		\hline
	\end{tabular}
	\label{tab:datasets_periodic}
\end{table}

\begin{figure}[!t]
    \centering
    \includegraphics[width=\linewidth]{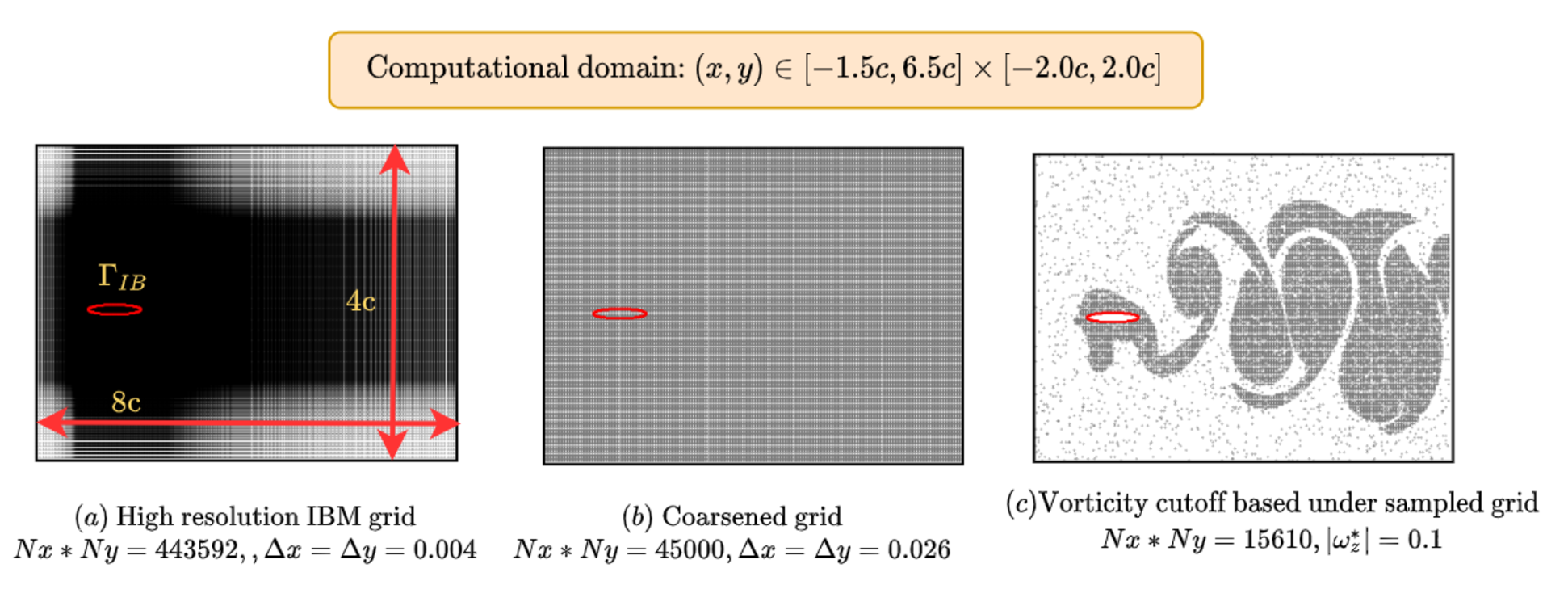}
    \caption{Computational domain and grids used for  training and testing for the quasi-periodic case; (a)  truncated high-resolution grid for IBM on which the models are  tested, (b) ~6x coarsened grid with ~10x lower number of data points than in the high-resolution case, (c ) vorticity cut off based undersampled grid. }
    \label{fig:qpgrids}
\end{figure}

   Coarsened simulation data, and subsequently vorticity cutoff based spatial sampling, with a sampling ratio of $S_{\omega_z}=5\%$,  were used for training, as was also considered in our earlier study~\cite{sundar2024physics}. For the periodic case, $|\omega^*| = 1$ was chosen. For the quasi-periodic case, where the Reynolds number was lower, a lower cutoff of $|\omega^*| = 0.1$ was chosen due to the possibility of relatively larger sizes of the flow-field vortices. This also led to a larger number of grid points sampled in the wake region, compared to the near-field region surrounding the moving body. A schematic of the truncated domain and a representative snapshot of the undersampled grids have been presented for the periodic and quasi-periodic  cases in figures~\ref{fig:pgrids} and \ref{fig:qpgrids}, respectively. 
 Note that the problem deals with a moving discontinuity in the domain and in that light, evaluating the accuracy of near-field velocity reconstruction and pressure recovery under different temporal complexities to identify the suitable  sequential learning variants is crucial. Towards that aim, the following databases were generated to investigate the efficacy of the sequential learning strategies: for the periodic case, the details of the dataset  are shown in table~\ref{tab:datasets_periodic}, and for the quasi-periodic case in table~\ref{tab:datasets_quasiperiodic}.

\begin{table}[htbp!]
	\centering
	\caption{Details of the training and testing databases for the domain considered in this study for the quasi-periodic flow. CI-*: coarse interpolated datasets; Ref-IBM-QS and Ref-ALE-QS: high-resolution testing data sets generated using an IBM and an ALE solver, respectively. The domain size considered here is $[-1.5c, 6.5c] \times [-2c, 2c].$ }
	\resizebox{\textwidth}{!}{\begin{tabular}{ccccccc}
		\hline
		\textbf{Data sets} & $N_x$ & $N_y$ & $N_t^{Bulk}$& $(\Delta t/T)_{Bulk}, (\Delta t/T)_{Phy}$ &$N_{Bulk}$ & $N_{Phy}$  \\
		\hline
            \multicolumn{7}{c}{Training datasets}\\
            \hline
            CI-QP81 & 300 & 150 & 81 & 0.125, 0.015625 & \textbf{3.594e06}&\\
            CI-S5-QP81 & - & - & 81 & 0.125, 0.015625 & \textbf{1.219e06}& \\
            CI-S5-QP21 & - & - & 21 & 0.5, 0.015625 & \textbf{3.219e05}& \\
            \hline
            \multicolumn{7}{c}{Testing datasets}\\
            \hline
            Ref-IBM-QP1 & 732 & 606 & 641 & 0.015625, - & 2.805e08 &-\\
		\hline
	\end{tabular}}
	\label{tab:datasets_quasiperiodic}
\end{table}

\section{Tackling temporal complexities in the  periodic case}\label{sec:periodicresults}
 The sequential learning strategies described in the previous section are first evaluated for the periodic flow case. As the flow-field is regular and repeats exactly in this case, it would be possible to identify the underlying data-independent deficiencies of the models. Specifically, two isolated training scenarios for the periodic  case study have been  considered: one with temporal sparsity of the velocity data over a short time domain, and the other with a long time domain but with temporally well resolved data. 
 We wish to determine the efficacy of the proposed sequential learning variants of MB-PINN under each case.  
Although MB-PINN is used here as the body position and its velocity information are known beforehand, these sequential learning methods can also be directly adopted for MB-IBM-PINN as was proposed in~\cite{sundar2024physics}, in case temporal domain complexities arise.

\subsection{Importance of physics loss, and residual collocation points}

\begin{figure}[!b]
    \centering
    \includegraphics[width=\linewidth]{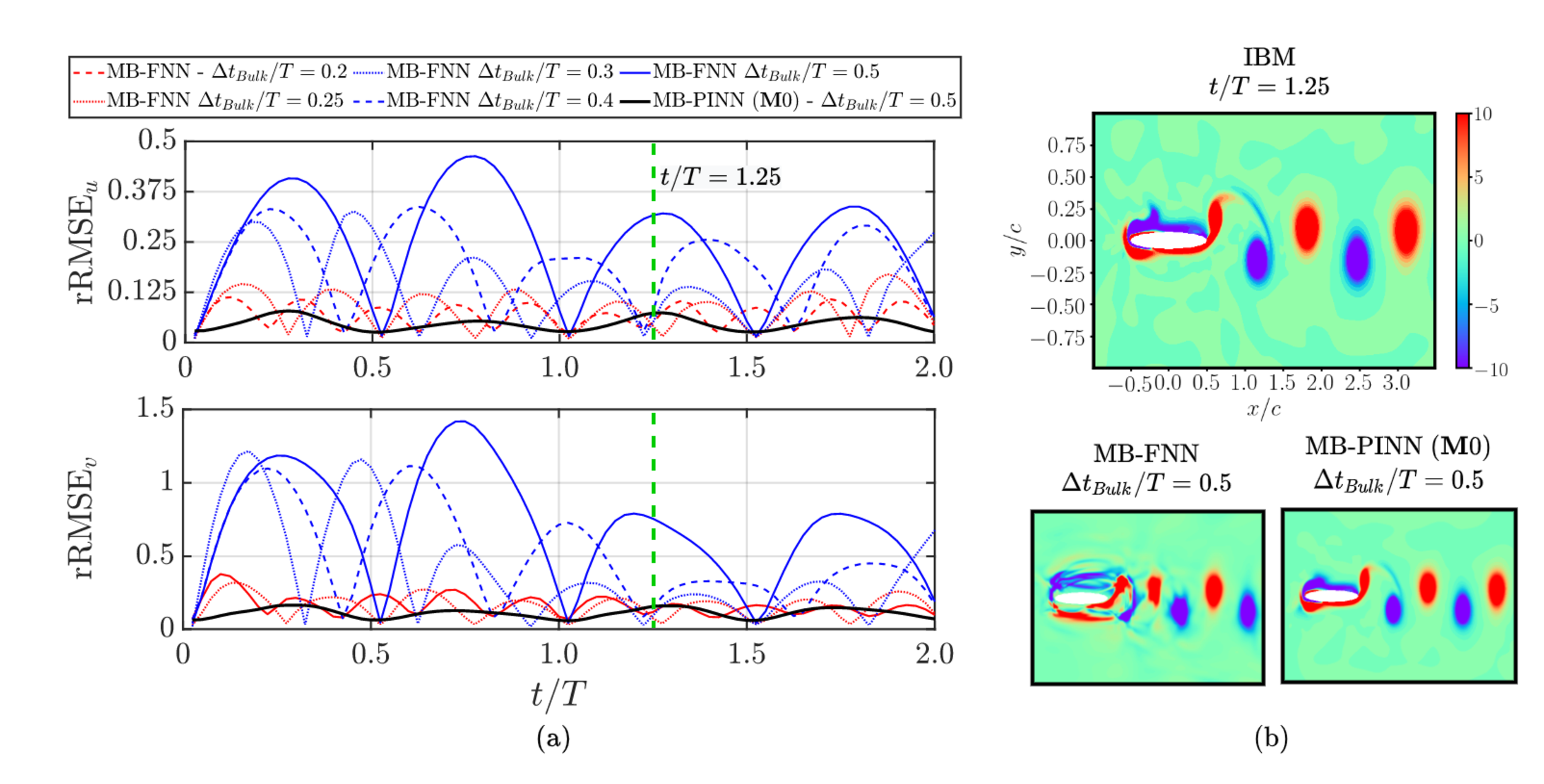}
    \caption{Comparison of standard MB-PINN (baseline, $\mathbf{M}_0$) with purely data-driven MB-FNN for velocity reconstruction at different $\Delta t_{Bulk} /T \in [0.2, 0.25, 0.3, 0.4, 0.5]$; (a) relative errors over time for velocity reconstruction, and,  (b) comparison of a representative vorticity snapshot obtained from (top) IBM ground truth, and (bottom) model predictions at $t/T = 1.25,$ where, a significant improvement in the near-field is observed with MB-PINN ($\mathbf{M0}$) as opposed to MB-FNN for $\Delta t_{Bulk}/T = 0.5.$}
    \label{fig:PIlossimportance}
\end{figure}

For an accurate reconstruction of temporal signals, Nyquist criterion suggests that the sampling rate should be at least twice the maximum frequency observed in the system,  that is, $\Delta t_{Bulk} / T \leq f_h/2f_{max}.$ Purely data-driven methods could fail when the temporal resolution does not adhere to the Nyquist criterion. In the context of flow-field reconstruction, especially with a moving body in the domain, temporal interpolation/reconstruction of the velocity field under temporal sparsity could be challenging. The interpolation errors would often be the highest in the vicinity of the moving body~\cite{sundar2024physics}. Hence, it is important to determine if there are any advantages in using the physics informed baseline MB-PINN ($\mathbf{M0}$) model for temporal interpolation, over its purely data-driven standard feed-forward neural network (MB-FNN) counterpart used in ~\cite{sundar2024physics}. To begin with, MB-FNN~\cite{sundar2024physics} would require a large amount of data to perform satisfactorily. Given that PINNs are expected to work well under data-sparsity conditions, the need for a physics informed loss is demonstrated by comparing the spatio-temporal velocity reconstruction through MB-FNN and $\mathbf{M0}$ model of MB-PINN under different temporal sparsity levels ($\Delta t_{Bulk}/T \in [0.2, 0.25, 0.3, 0.4, 0.5]$). In most cases, one can acquire the position and the velocity of the body  with high resolution through certain  means, but it might be difficult to obtain the flow-field data. Hence, frequency content of the flow from temporally sparse data cannot be estimated accurately {\it a priori} and hence the Nyquist criterion cannot be defined exactly for the PINN models. However, it is known that the vortex shedding frequency would most often be twice the plunging frequency~\cite{majumdar2020capturing}. Hence, a proxy Nyquist criterion can still be defined based on the plunging frequency as, $\Delta t_{Bulk}/T \leq f_h/2f_{max} = f_h/4f_h = 0.25$.  
Note that the predictions are made at a temporal resolution of $\Delta t/T = 0.025$ which is atleast $8$ times finer depending on the $\Delta t_{Bulk}/T$ mentioned  above. 

The relative reconstruction errors in time (figure~\ref{fig:PIlossimportance}(a)) for purely data driven MB-FNN across the temporal sparsity levels were  notably worse than $\mathbf{M0}$, trained with $\Delta t_{Bulk} = 0.5$ not satisfying the Nyquist criterion. Moreover, the near-field was distorted for the MB-FNN interpolated vorticity contour at a test time stamp, whereas, $\mathbf{M0}$ outperformed in this regard (see figure~\ref{fig:PIlossimportance}(b)). This clearly shows that the physics informed approach is capable of significantly improving the spatio-temporal interpolation in the limited data scenario.
A detailed comparison of the average error metrics of velocity reconstruction has been presented in table~\ref{tab:physlossimportance} for $\Delta t_{Bulk}/T = 0.5$.  

\begin{table}[!t]
\centering
\label{tab:detailedaccuracy_hn100hl10_stdpinn}
\caption{Comparison of the accuracy of  velocity component ($v$) reconstructed for the periodic case by  linear interpolation (LI),  purely data-driven feedforward neural network (MB-FNN), and  standard MB-PINN ($\mathbf{M0}$) when trained with varying levels of temporal sparsity without satisfying the Nyquist criterion ($\Delta t/T \leq 1/(2f_{max}) = 0.25$). The errors are evaluated over two plunging cycles.}
\label{tab:physlossimportance}
\resizebox{\textwidth}{!}{\begin{tabular}{ccccccc}
	\hline 
	\textbf{Model} &$\Delta t_{Bulk}/T$ &  \textbf{Network size} & \textbf{RMSE} & \textbf{MAE} & $R^2$ & \textbf{rRMSE}(in \%)\\ 
    \hline
        \textbf{LI} &0.5 &-&5.17e-01&3.03e-01&-1.01e-01&104.96\\
        \textbf{MB-FNN}& 0.5 &$100 \times 10$ & 3.56e-02&	1.87e-02&	3.77e-01&	69.87 \\ 
        \textbf{$\mathbf{M0}$} & 0.5 &$100 \times 10$ & 5.62e-02&	2.38e-02&	9.87e-01&	\textbf{11.07} \\ 
    \hline
\end{tabular} }
\end{table}

\begin{figure}[!t]
    \centering
    \includegraphics[width=0.8\linewidth]{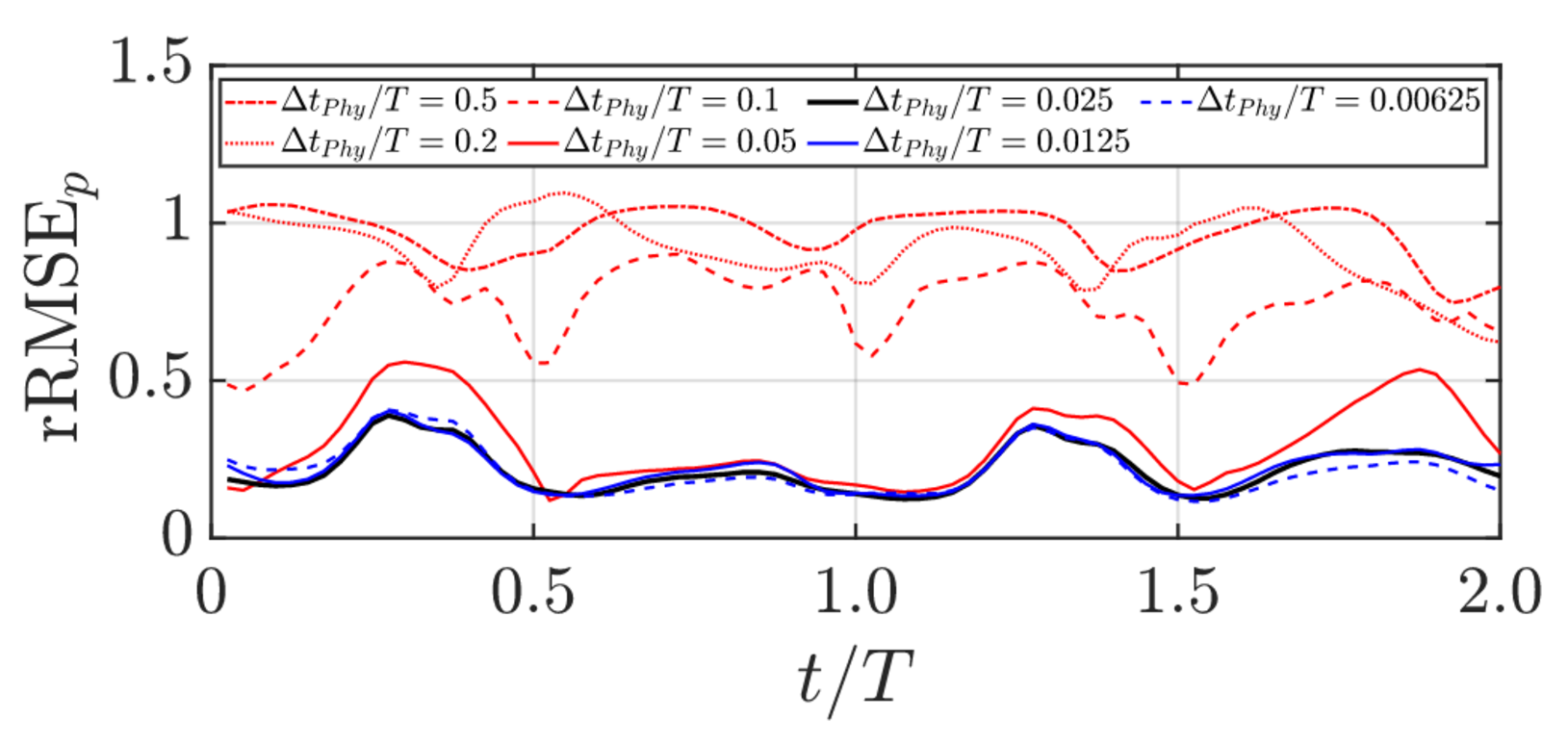}
    \caption{Comparison of the time behavior of relative RMSE  for  pressure recovery with MB-PINN ($\mathbf{M0}$) models trained over two plunging  periods for the periodic case. Effect of different temporal resolutions of residual collocation points,   $\Delta t _{Phy}/T \in [0.5, 0.2, 0.1, 0.05, 0.025, 0.0125, 0.00625]$ is shown. The data snapshots are available at a temporal resolution of $\Delta t_{Bulk}/T = 0.5.$ As temporal resolution increases ($\Delta t_{Phy}/T$ decreases), the pressure recovery improves overall.}
    \label{fig:DeltaTResImportance}
\end{figure}

With the help of its physics-informed approach in $\mathbf{M0}$,  it is also possible to recover pressure as a hidden variable (unlike MB-FNN). But the temporal resolution of the residual collocation points ($\Delta t_{Phy}/T$) could also play an important role in improving the accuracy of pressure recovery as seen in figure~\ref{fig:DeltaTResImportance}.
Notably, in figure~\ref{fig:DeltaTResImportance} at very coarse $\Delta t_{Phy}/T \in [0.5, 0.2, 0.1],$ the errors were above 50\% on average, indicating   a poor pressure recovery. Whereas, there was a significant improvement in the  relative error at $\Delta t_{Phy}/T = 0.05$, which then plateaued at $\Delta t_{Phy}/T = 0.025$ with  insignificant improvements at finer temporal resolutions. For practical reasons, having a coarser temporal resolution for residual collocation points and bulk data points, is computationally efficient. This is because, with coarser resolution the overall size of the entire dataset is reduced. Thus, during training, given a fixed mini-batch size, the model can see more number of epochs of the data at coarser resolutions. Therefore, a favorable trade-off needs to be achieved between accuracy and computational efficiency while selecting the temporal resolution. In this case, $\Delta t_{Phy} \in [0.0125, 0.025]$ was reasonable as indicated in figure~\ref{fig:DeltaTResImportance}. For more information on the effect of temporal resolution of residual collocation points, see ~\ref{sec:reseffect}, specifically, tables~\ref{tab:resaccuracyv}~and~\ref{tab:resaccuracyp}. 

\subsection{Efficacy of models under temporal sparsity}
It is evident that with physics-informed loss, and appropriate temporal resolution of residual collocation points, both spatio-temporal interpolation of velocity data, and pressure recovery are much better than the pure data-driven counterpart. Recent works~\cite{penwarden2023unified, mattey2022novel, wang2022respecting} suggest sequential learning methods result in improved performance due to gradual increase, or reduced temporal domain size (see section 2 for an  explanation on the classification of the methods and the different ways in which they reduce the problem complexity) as compared to the simple standard learning. These investigations mostly relied on the ability to solve forward problems for some canonical PDEs. For most practical applications, traditional physics based solvers are often triumphant in forward problems, owing to their computational efficiency. It is only for inverse problems PINNs are powerful, owing to their flexiblity and ability to work with limited data. In addition, in  most cases, some form of observational/simulation data are  available albeit not of desired level of spatio-temporal resolution and quality. This could be due to computational/memory constraints, which inhibit experimentalists and CFD engineers alike to deal with truncated spatial domains and coarse temporal resolutions. When data becomes limited in space-time, PINNs could prove to be useful for pressure recovery as a hidden variable. However, the standard learning based $\mathbf{M0}$ can be hard to train on temporally sparse data, collected over long time domains especially for moving body flow-fields. In such a  scenario, it is important to study the efficacy of sequential learning over standard learning methods.
For this purpose, the baseline, Class I ($\mathbf{M1}$), and Class II ($\mathbf{M2}$) models were first evaluated under temporal sparsity for velocity reconstruction and pressure recovery over a short time domain $t/T\in [0,2]$ (two plunging time periods); this was  with  $\Delta t_{Bulk}/T =0.5$ that did not meet the Nyquist criterion  ($\Delta t_{Bulk}/T < 0.25$). 
\begin{figure}[!t]
    \centering
    \includegraphics[clip=True, width=0.9\linewidth]{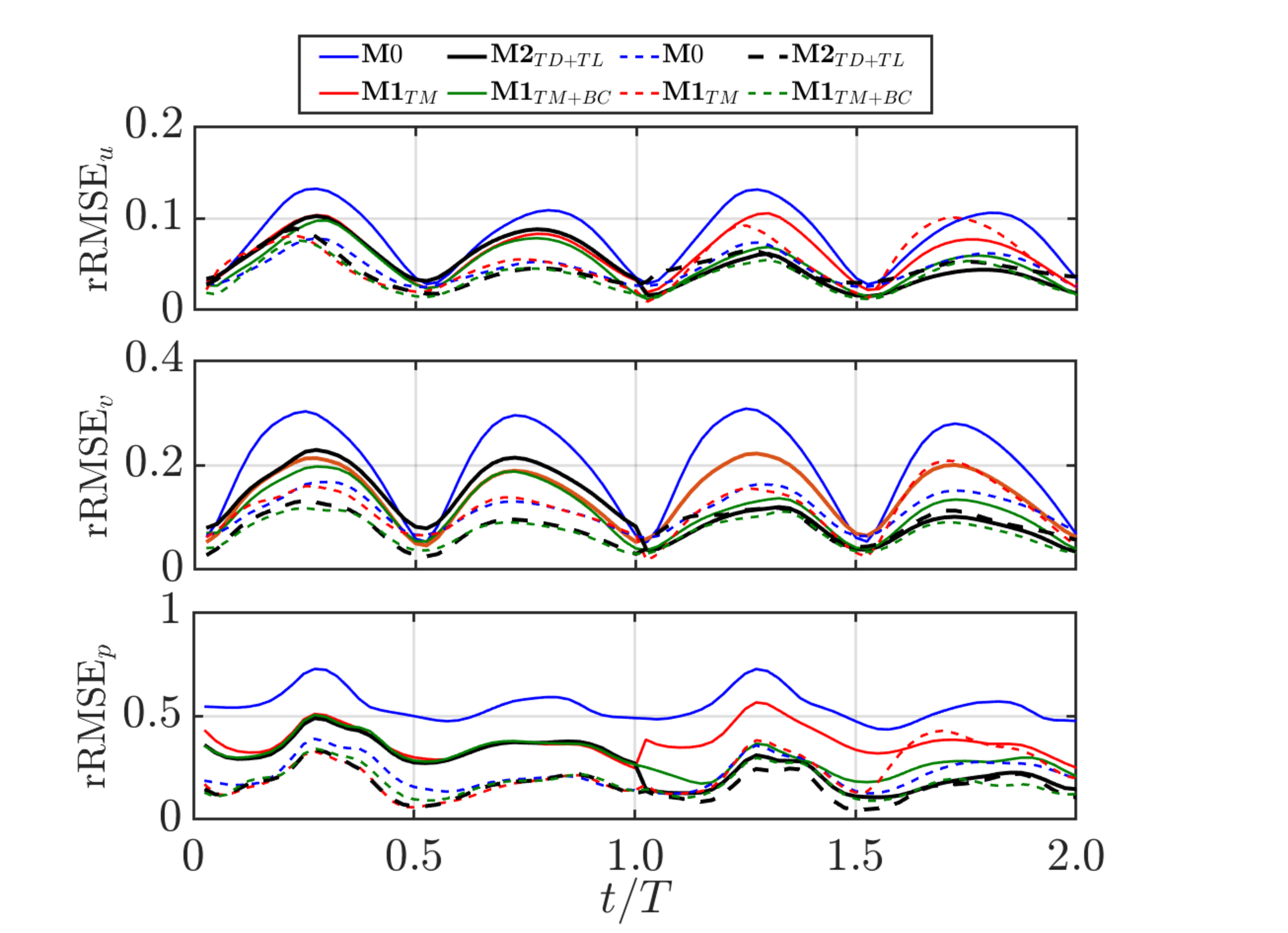}
    \caption{Relative RMSE over different times for (top-middle) reconstructed velocity, and (bottom) recovered pressure fields compared for different standard and sequential learning model variants for the periodic case. Solid lines represent shallow models with $l = 5$ hidden layers, dashed lines represent deeper models with $l = 10$ layers.}
    \label{fig:Psparsityrelerr}
\end{figure}

Based on the pressure recovery studies carried out using $\mathbf{M0}$ in ~\ref{sec:sparsityeffect}, (see Appendix table~\ref{tab:detailedaccuracy_pressurerecovery_stdpinn}), the networks of size ($100\times10$) and  ($100\times 5$) were chosen as  potential candidates for evaluation, ($50\times5$) could not yield a good pressure recovery which could be due to lack of model expressivity. 

The relative errors in time (see figure~\ref{fig:Psparsityrelerr}) for the reconstructed velocity and recovered pressure fields indicate the highest average errors for time stamps that are equidistant from the previous and next available data snapshots. It is evident that by design, in class I methods, the errors accumulate over the later time stamps ($\mathbf{\mathbf{M1}}_{TM}$), or, over the earlier time stamps ($\mathbf{\mathbf{M1}}_{TM+BC}$), respectively which is undesirable. In such cases class II models, by design, are triumphant over the standard MB-PINN. Since the kinematics is periodic and the time domain decomposition is such that initial position for both time segments align exactly, transfer learning was seen to be beneficial over the variants that see the problem complexity growing gradually. 
\begin{figure}[!htbp]
    \centering
    \includegraphics[width=\linewidth]{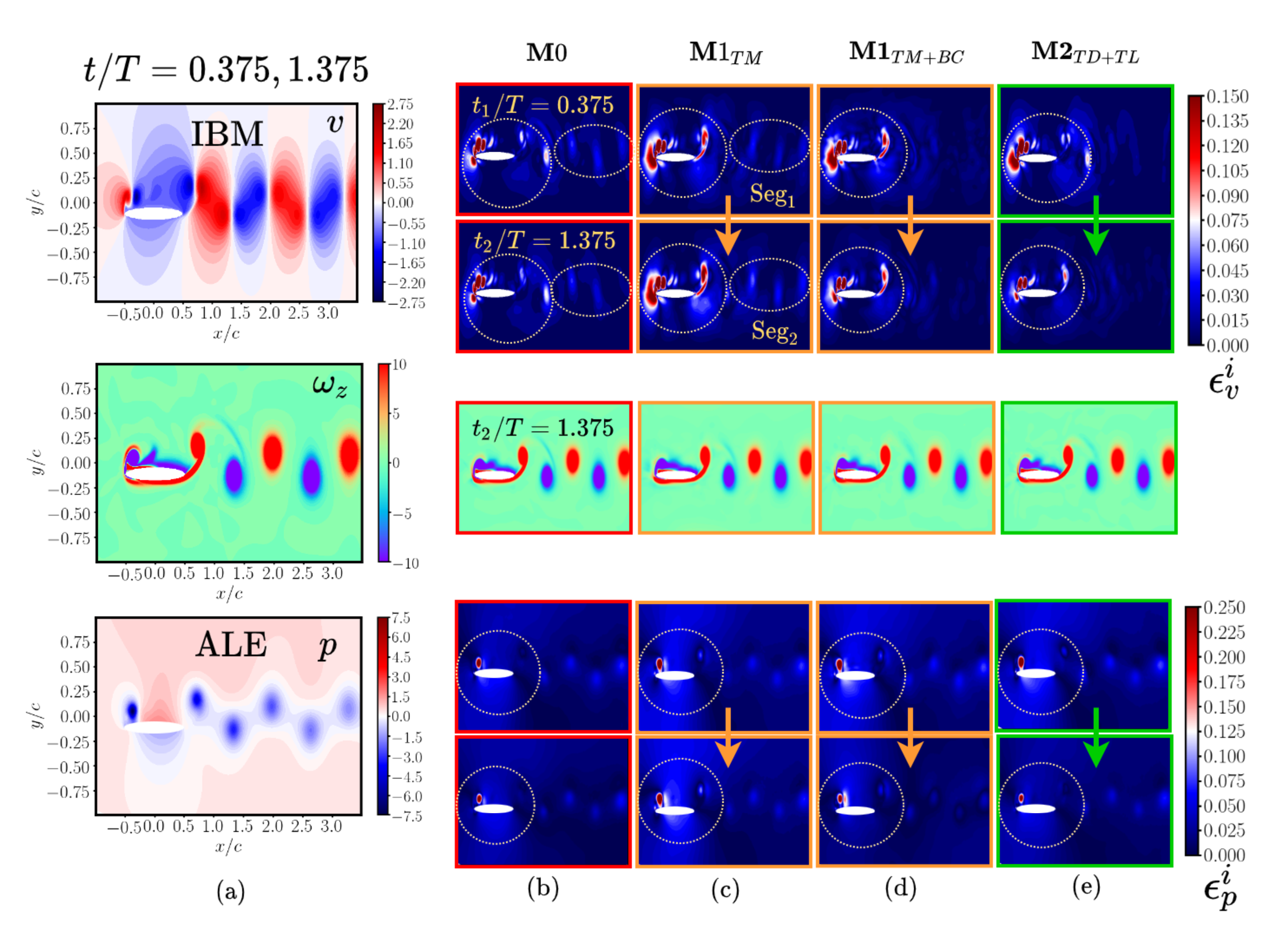}
    \caption{Comparison of point-wise maximum value normalized absolute errors in (top) velocity, and (bottom) pressure predictions and (middle) corresponding vorticity contours obtained from standard and sequential learning MB-PINN variants at representative test time stamps from first and last time segments at $\Delta t_{Bulk}/T = 0.5$ for the periodic case,}
    \label{fig:Psparsitycontours}
\end{figure}

The velocity, pressure and vorticity contours in figure~\ref{fig:Psparsitycontours} reveal that highest errors are observed consistently in the near-field region surrounding the moving body, across different model variants. Whereas, the far-field predictions are almost qualitatively indiscernible from the ground truth. Based on the averaged error metrics in tables~\ref{tab:detailedaccuracy_sparsity_v} and ~\ref{tab:detailedaccuracy_sparsity_p}, out of  Class I and Class II models, $\mathbf{M1}_{TM+BC}$, and $\mathbf{M2}_{TD+TL}$  were promising for handling temporally sparse short time domain datasets. With the benefit of transfer learning, class II variant ($\mathbf{M2}_{TD+TL}$) performed significantly better than $\mathbf{M1}_{TM+BC}$, especially when the network size was  smaller ($n = 100$ and  $l=5$), which is a desirable feature in terms of  computational efficiency.

\begin{table}[!htbp]
\centering
\caption{Comparison of the accuracy of  velocity component $v$  of the periodic case reconstructed by linear interpolation (LI), and the standard, sequential, and backward compatible MB-PINN models when trained with $\Delta t_{Bulk}/T = 0.5$ without satisfying the Nyquist criteria ($\Delta t_{Bulk}/T\leq 1/(2f_{max}) = 0.25$). The errors are evaluated over two plunging cycles.}
\resizebox{\textwidth}{!}{\begin{tabular}{cccccc}
	\hline 
	\textbf{Model} & \textbf{ Network size }& \textbf{RMSE} & \textbf{MAE} & $R^2$ & \textbf{rRMSE}(in \%)\\ 
        \hline 
        \textbf{LI}& -&5.17e-01&3.03e-01&-1.01e-01&104.96\\
        \hline 
	\multirow{2}{*}{\textbf{$\mathbf{M0}$}} & $100 \times 5$ & 1.02e-01&	4.77e-02&	9.53e-01&	20.13 \\
	&$100 \times 10$ & 5.62e-02&	2.38e-02&	9.87e-01&	11.07 \\ 
    \hline
	\multirow{2}{*}{$\mathbf{M1}_{TM}$} & $100 \times 5$ & 7.04e-02&	3.29e-02&	9.78e-01&	13.84 \\
        &$100 \times 10$ & 5.73e-02&	2.04e-02&	9.85e-01&	11.29 \\
        \hline
	\multirow{2}{*}{\textbf{$\mathbf{M1}_{TM+BC}$}} & $100 \times 5$ & 7.14e-02&	3.25e-02&	9.77e-01&	14.02 \\
        &$100 \times 10$ & 4.30e-02& 1.76e-02&	9.91e-01&	8.45 \\
        \hline
	\multirow{2}{*}{$\mathbf{M2}_{TD+TL}$} & $100 \times 5$ & 4.84e-02&	2.09e-02&	9.88e-01&	\textbf{9.49} \\
        & $100 \times 10$ & 4.23e-02&	1.52e-02&	9.91e-01&	\textbf{8.29} \\
    \hline
\end{tabular}}
\label{tab:detailedaccuracy_sparsity_v}
\end{table}

\begin{table}[!htbp]
\centering
\caption{Comparison of the pressure recovery accuracy of the standard, sequential, and backward compatible MB-PINN models for the periodic case, The models are trained with $\Delta t_{Bulk}/T = 0.5$ without  satisfying the Nyquist criteria ($\Delta t_{Bulk}/T\leq 1/(2f_{max}) = 0.25$). The errors are evaluated over two plunging cycles.}
\begin{tabular}{cccccc}
	\hline 
	 \textbf{Model} & \textbf{Network size} & \textbf{RMSE} & \textbf{MAE} & $R^2$ & \textbf{rRMSE}(in \%)\\ 
        \hline
	\multirow{2}{*}{\textbf{$\mathbf{M0}$}} &$100 \times 5$ & 7.22e-01&	4.87e-01&	6.95e-01&	54.71 \\ 
	&$100 \times 10$ & 2.82e-01&	1.77e-01&	9.47e-01&	21.74 \\ 
        \hline
	\multirow{2}{*}{$\mathbf{M1}_{TM}$} & $100 \times 5$ & 4.89e-01&	3.12e-01&	8.57e-01&	37.12 \\
        & $100 \times 10$ & 2.89e-01&	1.57e-01&	9.41e-01&	22.11 \\
         \hline 
	\multirow{2}{*}{$\mathbf{M1}_{TM+BC}$} & $100 \times 5$ & 4.63e-01&	2.96e-01&	8.72e-01&	35.23 \\
        & $100 \times 10$ & 2.48e-01&	1.48e-01&	9.57e-01&	\textbf{19.21} \\
  \hline 
	        \multirow{2}{*}{$\mathbf{M2}_{TD+TL}$} & $100 \times 5$ & 3.19e-01&	1.96e-01&	9.28e-01&	\textbf{24.37} \\
        &$100 \times 10$ & 2.60e-01 & 1.42e-01 & 9.50e-01 & 20.19\\
    \hline
\end{tabular}
\label{tab:detailedaccuracy_sparsity_p}
\end{table}

\subsection{Efficacy of models when trained over a long time domain}

The models were trained over a time domain encompassing 10 plunging time periods.
While there is sufficient temporal resolution, it is important to determine which variant is suitable especially for pressure recovery when trained over long-time domain data. Long-time domains might  pose training difficulties to PINNs  when there are strong flow-field gradients and a moving boundary, it is expected that sequential training variants will perform better as they reduce the problem complexity while keeping the models expressive. 

\begin{figure}[!htbp]
    \centering
    \includegraphics[width=0.95\linewidth]{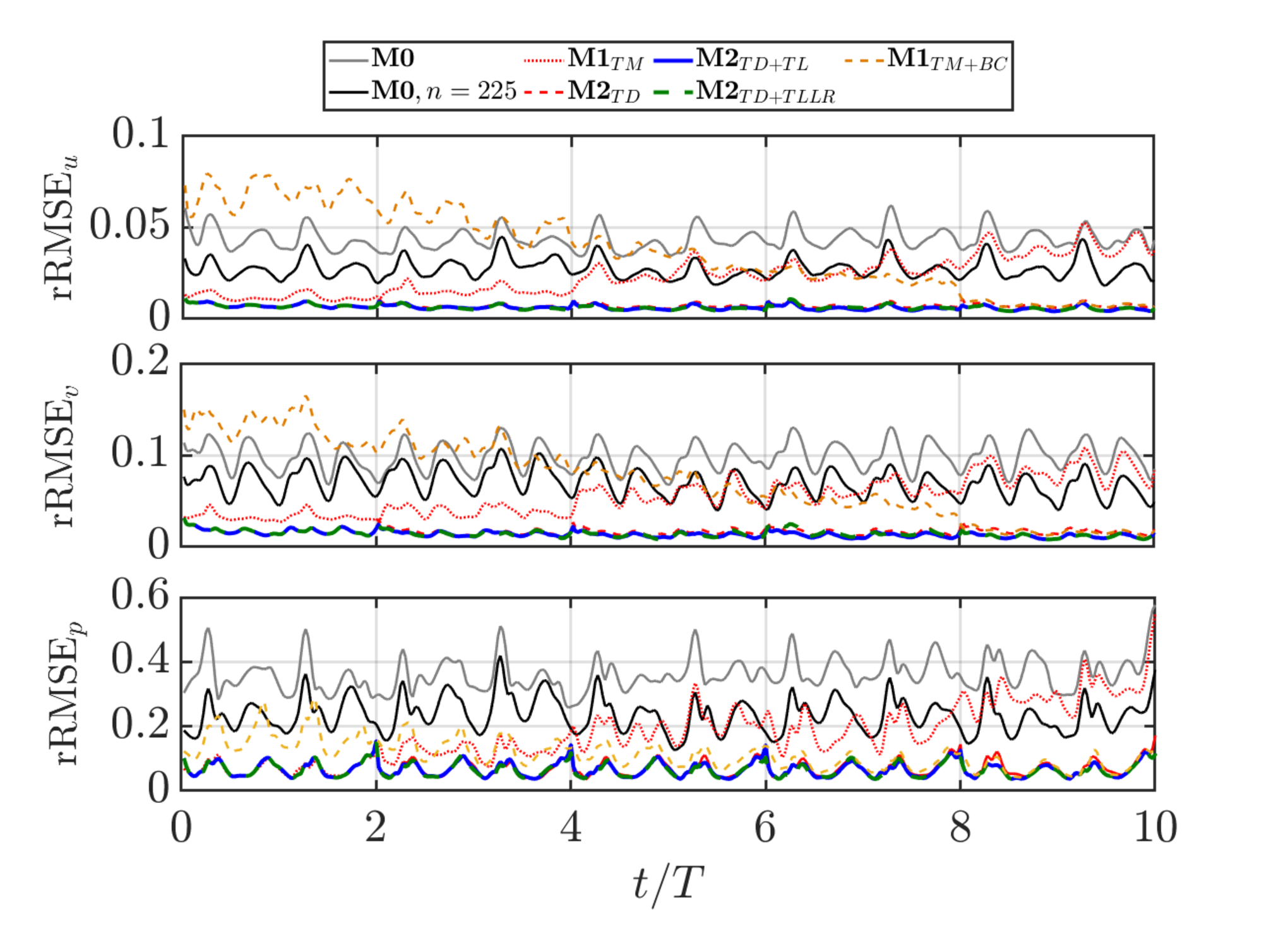}
    \caption{Comparison of snapshot-wise relative error in (a-b) velocity reconstruction and (c) pressure recovery obtained for the standard and sequential learning based MB-PINN models for the long time domain case study.}
    \label{fig:Plongtimerelerr}
\end{figure}

Sequential learning methods in the earlier subsections have so far been tested for sparsity, however, it is equally important to note that practical applications require these methods to be evaluated over long time domains. To isolate  the effects of  a long time domain usage, class I and II model variants were evaluated  with   $t/T\in [0,10],$ and $\Delta t_{Bulk}/T = 0.05$ for the periodic case. The choice of $\Delta t_{Bulk}/T$ ensured the Nyquist criteria was satisfied by a large margin. 

\begin{figure}[!t]
    \centering
    \includegraphics[width=\linewidth]{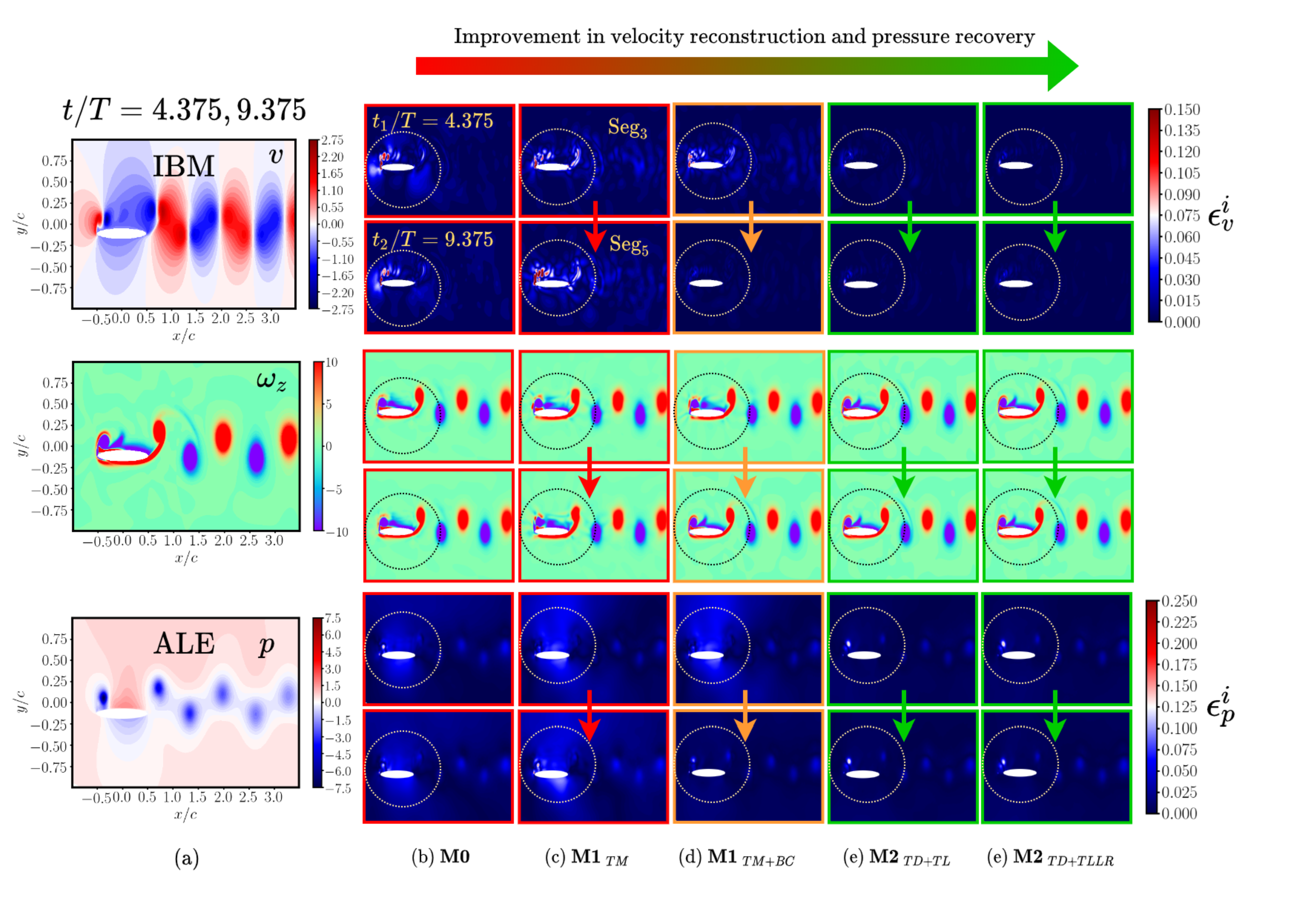}
    \caption{Comparison of (a) ground truth (top) $y$ velocity component $v$ pointwise error, (mid) vorticity, and (bottom) pressure pointwise error contours for predictions of (b) standard and (c-f) sequential learning based models across two different time domain segments}
    \label{fig:Plongtimecontours}
\end{figure}

Figure~\ref{fig:Plongtimerelerr} shows that the baseline $\mathbf{M0}$,  of size  $n = 100$ and  $l = 10$, struggled to reconstruct the velocity, and even more so to recover pressure over this long time domain. This was the case even when a five times larger network was considered ($n 
= 225$ vs $n = 100$ hidden neurons with a fixed $l = 10$ hidden layers.) This was a consequence of the increased time domain size, which  subsequently was not met with a similar increase in model expressivity, requiring  the problem to be broken into simpler parts and that the model should not be too large. Sequential learning methods are promising since they either increase the problem complexity gradually (Class I), or, reduce the problem and model complexity through temporal domain decomposition (Class II).

In the sparsity case study presented earlier, Class I model variants  performed relatively better than the $\mathbf{M0}$ models (see tables~\ref{tab:detailedaccuracy_sparsity_v} and \ref{tab:detailedaccuracy_sparsity_p}). But, the problem of  error accumulation still persisted in Class I models, which was more evident in the long time domain case. This has been very clearly revealed in the relative error plots in figure~\ref{fig:Plongtimerelerr}.  Here $\mathbf{M1}_{TM}$ can be seen accumulating  error over time, for the later  segments, while $\mathbf{M1}_{TM+BC}$ accumulating error over all the prior time segments. However, Class II models, $\mathbf{M2}_{TD}$, $\mathbf{M2}_{TD+TL},$ or the more computationally efficient $\mathbf{M2}_{TD+TLLR}$, performed better than Class I and $\mathbf{M0}$, overall. The $\mathbf{M2}_{TD+TLLR}$ variant involved training the subsequent time segments over only two learning rate decay stages, with the starting learning rate being $2e-04$ as opposed to $1e-03$ for the other variants. As a result,  $\mathbf{M2}_{TD+TLLR}$ required lower number of iterations ($2/3\times$), while still performing comparably to $\mathbf{M2}_{TD+TL}$ indicating a good computational efficiency.

This is confirmed in further qualitative inspection of the point-wise errors in velocity, vorticity and pressure as presented in figure~\ref{fig:Plongtimecontours}. Quantitatively, the averaged error metrics in tables~\ref{tab:detailedaccuracy_longtime_v} and \ref{tab:detailedaccuracy_longtime_p} also reveal that class II models are significantly better in performance.

\begin{table}[!htbp]
\centering
\caption{Long temporal domain: Comparison of accuracy details of reconstructed velocity component ($v$) by the standard, sequential, and backward compatible MB-PINNs when trained over  10 plunging cycles for the periodic case. Here, ($\Delta t_{Bulk}/T = \Delta t_{Phy}/T = 1/20$). The sequential and backward compatible variants are trained over 5 segments of 2 plunging cycles each.}
\label{tab:detailedaccuracy_longtime_v}
\resizebox{\textwidth}{!}{\begin{tabular}{cccccc}
	\hline 
	Model & Network size & \textbf{RMSE} & \textbf{MAE} & $R^2$ & \textbf{rRMSE}(in \%)\\ 
        \hline
        $\mathbf{M0}$ & $100\times 10$ &  4.96e-02 &	2.5e-02&	9.86e-01 &	9.86 \\ 
	$\mathbf{M0}$ &$225\times 10$& 3.51e-02&	1.68e-02&	9.95e-01&	6.97 \\ 
	$\mathbf{M1}_{TM}$ &$100\times 10$ & 2.75e-02&	1.42e-02&	9.96e-01&	5.49 \\ 
    $\mathbf{M1}_{TM+BC}$ &$100\times 10$ & 3.73e-01&	1.88e-02&	9.93e-01&	7.45\\ 
        $\mathbf{M2}_{TD}$ &$100\times 10$ & \textbf{8.39e-03}&	\textbf{4.67e-03}&	\textbf{9.99e-01}&	\textbf{1.68} \\
        $\mathbf{M2}_{TD+TL}$ &$100\times 10$ & \textbf{7.00e-03}&	\textbf{3.87e-03}&	\textbf{9.99e-01}&	\textbf{1.41} \\
        $\mathbf{M2}_{TD + TLLR}$ &$100\times 10$ & \textbf{7.59e-03}&	\textbf{3.99e-03}&	\textbf{9.99e-01}&	\textbf{1.52} \\
    \hline
\end{tabular}} 
\end{table}

\begin{table}[!htbp]
\centering
\caption{Long temporal domain: Comparison of the accuracy  of recovered pressure ($p$) by the standard, sequential, and backward compatible MB-PINNs when trained over  10 plunging cycles for the periodic case. Here, ($\Delta t_{Bulk}/T = \Delta t_{Phy}/T = 1/20$). The sequential and backward compatible variants are trained over 5 segments of 2 plunging cycles each. }
\label{tab:detailedaccuracy_longtime_p}
\resizebox{\textwidth}{!}{\begin{tabular}{cccccc}
	\hline 
	Model & Network size & \textbf{RMSE} & \textbf{MAE} & $R^2$ & \textbf{rRMSE}(in \%)\\ 
        \hline
        $\mathbf{M0}$ & $100\times 10$ &  4.79e-01 &	3.43e-01&	8.66e-01 &	36.25 \\ 
	$\mathbf{M0}$ &$225\times 10$& 3.09e-01&	2.16e-01&	9.41e-01&	23.62 \\ 
	$\mathbf{M1}_{TM}$ &$100\times 10$ & 2.31e-01&	1.65e-01&	9.59e-01&	17.94 \\ 
            $\mathbf{M1}_{TM+BC}$ &$100\times 10$ & 1.45e-01&	8.14e-02&	9.85e-01&	11.39 \\
        $\mathbf{M2}_{TD}$ &$100\times 10$ & \textbf{8.81e-03}&	\textbf{5.58e-03}&	\textbf{9.94e-01}&	\textbf{7.05} \\
        $\mathbf{M2}_{TD+TL}$ &$100\times 10$ & \textbf{8.36e-03}&	\textbf{5.11e-02}&	\textbf{9.95e-01}&	\textbf{6.66} \\
        $\mathbf{M2}_{TD + TLLR}$ &$100\times 10$ & \textbf{8.25e-03}&	\textbf{4.99e-03}&	\textbf{9.95e-01}&	\textbf{6.60} \\
    \hline
\end{tabular}} 
\end{table}

Overall, it was observed in the periodic flow case that $\mathbf{M1}_{TM}$ and $\mathbf{M1}_{TM+BC}$ are conclusively less efficient, in case of temporal sparsity, or, training over long time domain. This is because they train over a gradually increasing time domain. As a result, a larger network might be required to accommodate the entire time domain, towards the end of the training stage (when training over the last time segment). Here the major impediment is the error accumulation problem of Class I models,  which can be avoided with Class II ones. 
Moreover, the transfer learning based variants of Class II allow the overall problem size to be broken down, so that smaller networks could be trained to obtain a better accuracy, compared to $\mathbf{M0}$ or Class I models. 

In both individual  scenarios, the case of temporally sparse data, and the long time domain data well resolved in time, the time domain decomposition based Class II models with transfer learning ($\mathbf{M2}_{TD+TL}, \mathbf{M2}_{TD+TLLR}$)  were successful. Specifically, the $\mathbf{M2}_{TD+TLLR}$ model was more computationally efficient.\\ In the next section, these models will be used to reconstruct a quasi-periodic flow case. In addition, a simple but efficient preferential spatio-temporal sampling will also be proposed to improve the reconstructions in a data-efficient manner. 

\section{Tackling quasi-periodic flow with preferential spatio-temporal sampling}\label{sec:quasiperiodicresults}

Quasi-periodicity is encountered in certain kinematic regime  in the context of unsteady flows past flapping wings due to the aperiodic behaviour of the near-field vortex structures,  such regimes may enhance the aerdynamic load generation~\cite{bose2018transient, majumdar2020capturing, majumdar2022transition}. Enhanced load generation might also lead to a higher performance efficiency~\cite{majumdar2022transition}, which could be significant in the  design of  bio-mimetic devices. Quasi-periodic flow-field around a flapping body may  involves minor variations in the position/organization of the vortices though it may look seemingly regular. However, one of the most important characteristics of the dynamics is the presence of  incommensurate frequencies (and there combinations) in the system during quasi-periodicity.
Thus, quasi-periodic flow serves as a good example of temporal complexity that involves a rich temporal spectrum, to evaluate the sequential learning based models proposed. 

In the backdrop of our earlier discussion in section~\ref{sec:periodicresults}, for the quasi-periodic test case only $\mathbf{M2}_{TD+TL}$ and $\mathbf{M2}_{TD+TLLR}$ was evaluated against the $\mathbf{M0}$ model. For multi-scale problems,  earlier studies have suggested Fourier feature embeddings~\cite{wang2021eigenvector}, however, when the rich frequency content cannot be estimated {\it a priori} with temporally sparse data snapshots, such embeddings are not feasible without knowing  the exact frequency scales. Even when a sufficiently large frequency scale was chosen to initialize the learnable Fourier feature layers, preliminary results indicated that the predictions could not improve significantly. Hence, Fourier feature embeddings were not employed in this study. 

\subsection{Importance of temporal resolution in the near-field}

As mentioned previously, undersampling with the help of a vorticity  cut-off  was proposed in our earlier study~\cite{sundar2024physics} to improve data efficiency, while maintaining accuracy. This cut-off-based undersampling gives uniform weight to all the grid points, with data where the absolute vorticity $|\omega|$ is greater than a cut-off value  $|\omega^*|.$ 
For the quasi-periodic case study, a cut-off value  $|\omega^*| = 0.1$ was chosen and a baseline domain $\Omega^r_1$ was chosen $\Omega^r_1: [-1.5c,6.5c]\times[-2c,2c]$ to cover two vortex couples in the wake. In this case the domain is  larger than the periodic case study,   due to larger vortices involved (an effect of  lower Reynolds number  and larger plunging amplitude
). 
\begin{figure}[!htbp]
    \centering
    \includegraphics[clip=True, trim = {0 3cm 0 0}, width=\linewidth]{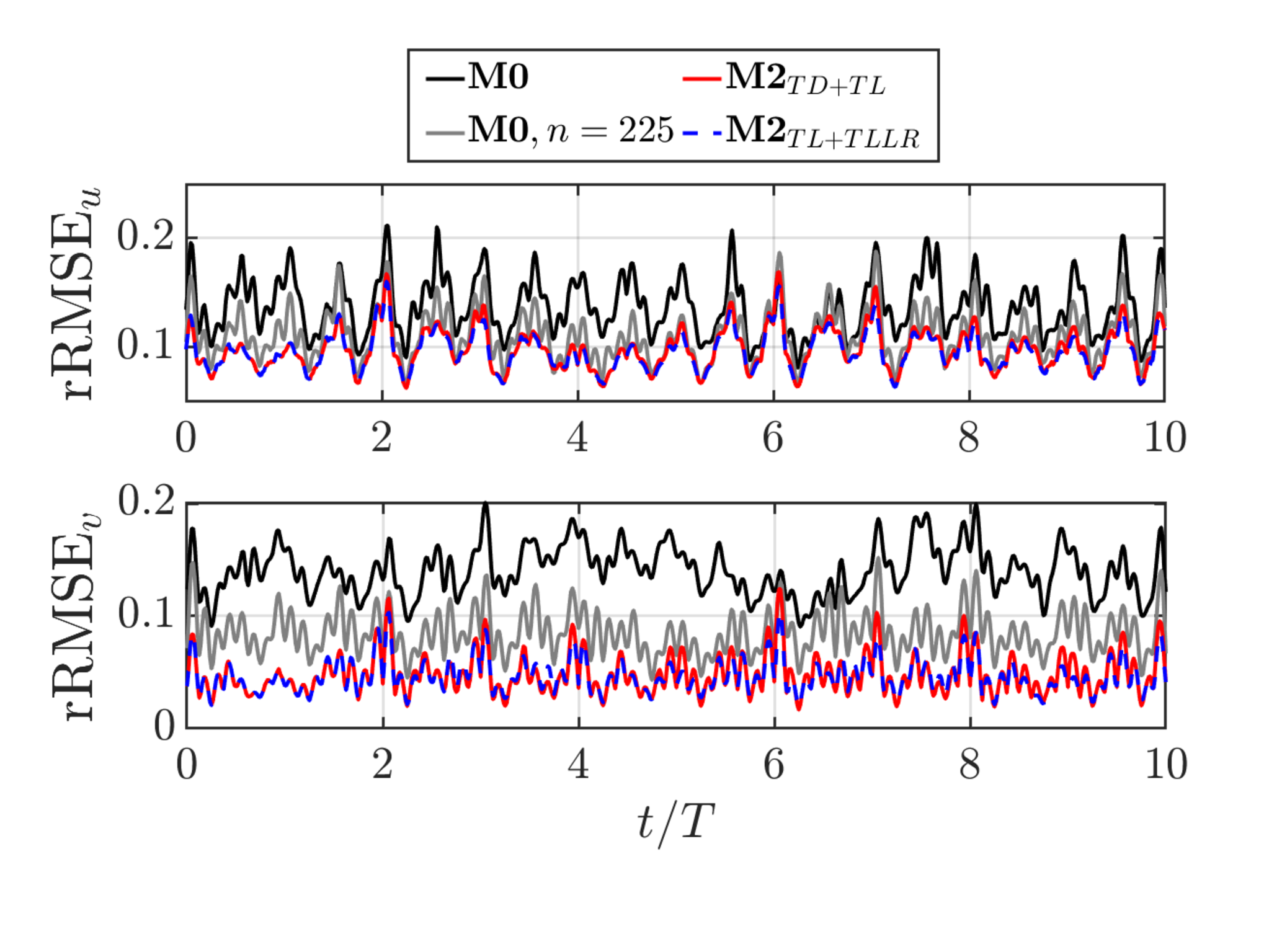}
    \caption{Comparison of relative error in time obtained for (top) $x-$ velocity component $u,$  and (bottom) $y-$ velocity component $v,$  reconstruction for $\Delta t_{Bulk}/T = 0.125.$ Unless otherwise specified, the networks  consist of 10 hidden layers with $n = 100$ neurons per hidden layer.}
    \label{fig:relerr_Re300kh1p65}
\end{figure}

\begin{figure}[!htbp]
    \centering
    \includegraphics[width=\linewidth]{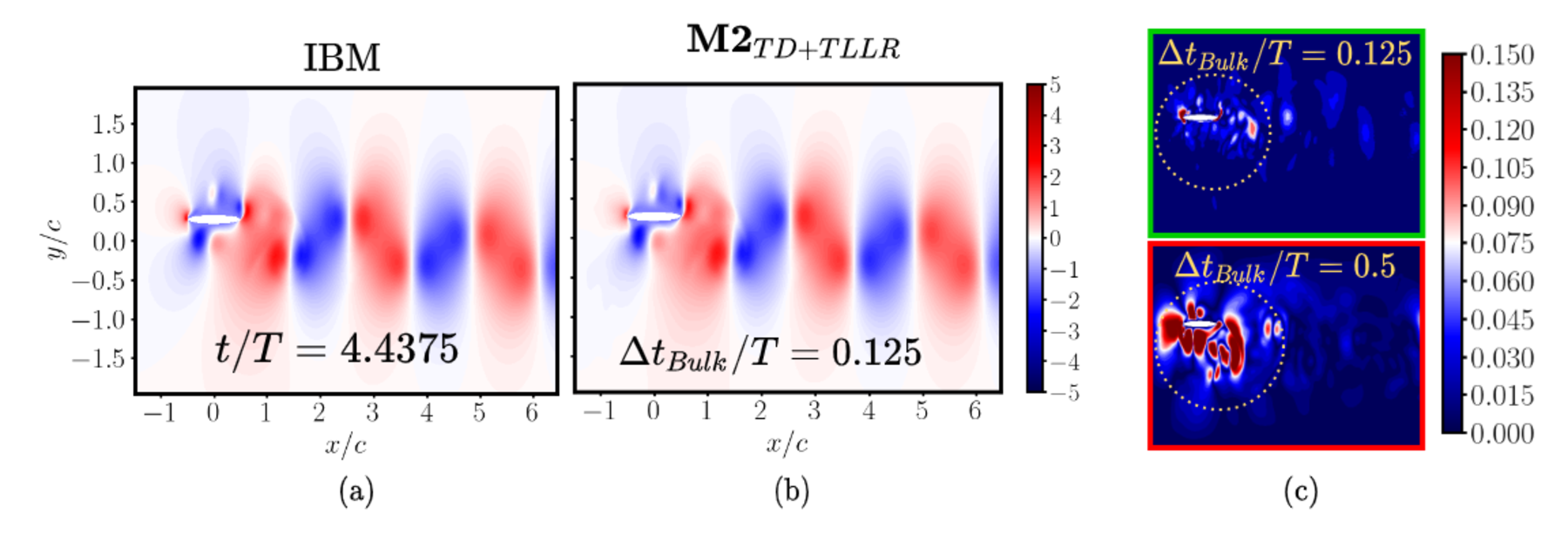}
    \caption{Comparison of (a) ground truth $y-$velocity and pressure data with (b) $\mathbf{M2}_{TD+TLLR}$ reconstruction at $\Delta t_{Bulk}/T = 0.125.$ The pointwise errors in (c) are compared for $\mathbf{M2}_{TD+TLLR}$ cases with $\Delta t_{Bulk}/T = 0.125, 0.5,$ respectively. Higher the sparsity the near-field reconstructions suffer the most while the wake is still reasonably reconstructed due to smoothness of the solution and no moving boundary in that region.}
    \label{fig:nearvsfarfield}
\end{figure}

The $\mathbf{M0}$, $\mathbf{M2}_{TD+TL}$, and $\mathbf{M2}_{TD+TLLR}$ models were trained over a long time domain for different $\Delta t_{Bulk}/T\in[0.125, 0.5],$ respectively. For $\Delta t_{Bulk}/T = 0.125,$ the relative RMSE (rRMSE) over time in figure~\ref{fig:relerr_Re300kh1p65} indicated that both $\mathbf{M2}_{TD+TL}$, and $\mathbf{M2}_{TD+TLLR}$ are equivalent and infact better than the baseline $\mathbf{M0}$ trained with different network sizes. Recall that, as per the Nyquist sampling criterion, $\Delta t_{Bulk}/T \leq 0.25.$ The rRMSE worsened when this criterion was not satisfied (see table~\ref{tab:detailedaccuracy_nearfar_v}), with the best performance at $\Delta t_{Bulk}/T = 0.125$. A closer inspection of the contours in figure~\ref{fig:nearvsfarfield} reveals the pointwise errors to be highest in the near-field region surrounding the moving boundary, and significantly lower errors in the wake region for both $\Delta t_{Bulk}/T$ considered. This points to the ability of sequentially trained MB-PINNs to reconstruct the wake even when $\Delta t_{Bulk}/T = 0.5$ does not satisfy the Nyquist criteria. However, accurate reconstruction of the velocity field and recovery of pressure in the near-field and moving body are crucial for the computation of the aerodynamic loads.

\begin{table}[!t]
\centering
\caption{Comparison of the accuracy  of  velocity component $v$  reconstructed by  standard MB-PINN and variants of seq-MB-PINN  when trained with varying levels of temporal sparsity over 10 plunging cycles (Quasi-periodic case) considering the spatial domain $\Omega^r_1$ as previously described, and $\lambda_{IB} = 1$.}
\resizebox{\textwidth}{!}{\begin{tabular}{cccccc}
	\hline 
	$\Delta t_{Bulk}/T$ &  \textbf{Network size} & \textbf{RMSE} & \textbf{MAE} & $R^2$ & \textbf{rRMSE}(in \%)\\ 
        \hline 
        \multicolumn{6}{c}{\textbf{$\mathbf{M0}$}} \\
        \hline
        0.125 &$100 \times 10$ &  1.27e-01&	7.33e-02&	9.79e-01&	14.03 \\ 
        0.125 &$225 \times 10$ &  7.72e-02&	4.02e-02&	9.92e-01&	8.52 \\ 
        \hline 
        \multicolumn{6}{c}{\textbf{$\mathbf{M2}_{TD+TL}$}} \\
        \hline
        0.125 &$100 \times 10$ &  6.34e-02&	3.13e-02&	9.95e-01&	\textbf{7.02 }\\ 
	0.5 & $100 \times 5$ & 2.14e-01&	1.03e-01&	9.38e-01&	23.67 \\ 
        \hline
        \multicolumn{6}{c}{$\mathbf{M2}_{TD+TLLR}$ } \\
        \hline
        0.125 &$100 \times 10$ &  6.49e-02&	3.09e-02&	9.94e-01&	\textbf{7.18} \\ 
	0.5 &$100 \times 5$ & 2.21e-01&	1.11e-01&	9.33e-01&	24.34 \\ 
    \hline
\end{tabular}}
\label{tab:detailedaccuracy_nearfar_v}
\end{table}

The near-field is affected possibly due to a relatively lower proportion of grid points in this region compared to the wake region. This results in PINNs automatically getting more updates from the wake region during training.  Moreover, since the flow-field is smoother in the wake region without discontinuity like the moving body, the far-field data reconstruction and pressure recovery are more robust during temporal sparsity, long time domains, and spatially sparse data. Further, the moving body region faces three competing objectives in the form of physics, bulk data, and the no-slip-boundary-condition loss components. Whereas, the wake region has only two competing objectives, the bulk-data and the physics loss components, respectively. Thus, multiple competing objectives, lack of sufficient data, and strong flow-field gradients make the training  difficult in the near-field region.

To tackle these challenges and improving the pressure recovery and subsequent load reconstruction, a preferential spatio-temporal sampling strategy (PVS), in addition to vorticity cut-off sampling has been  proposed in the next section.

 \begin{figure}[!b]
    \centering
    \includegraphics[width=\linewidth]{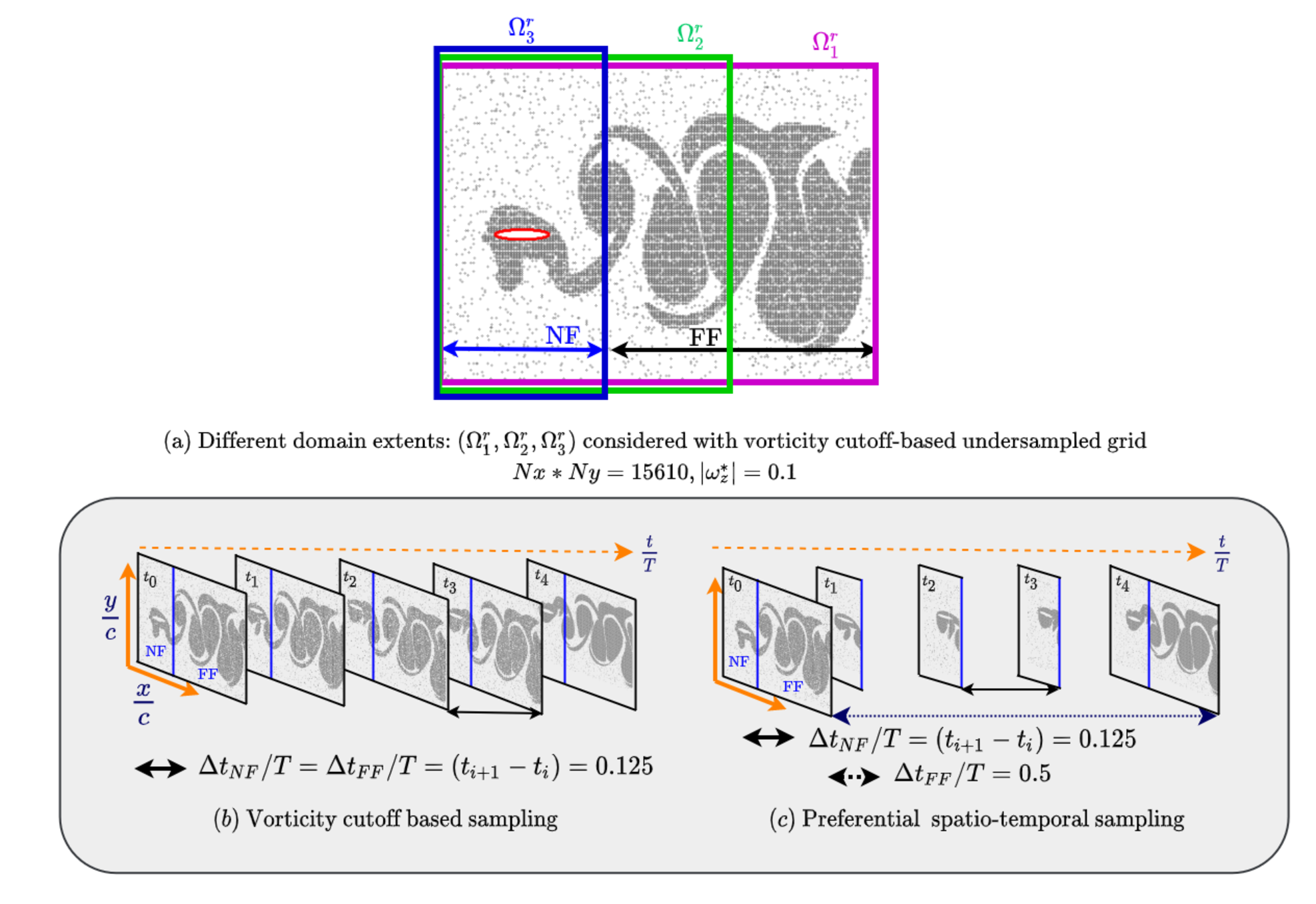}
    \caption{Schematic comparing the vorticity-based sampling (b) and preferential spatiotemporal sampling - combined with the vorticity-based sampling -  in the near and far-field regions (c).}
    \label{fig:prefsampling}
\end{figure}

\subsection{Preferential spatiotemporal sampling}

To improve the overall velocity reconstruction and pressure recovery, the preferential spatiotemporal sampling strategy (PVS) is proposed. In statistics, stratified sampling involves sampling from a population of data stratified or partitioned into sub-populations. The proposed PVS approach has similarities with such stratified sampling strategies. Here, the whole domain is first stratified/partitioned into near-field and wake region sub-domains, allowing flexibility to sample preferentially in space and time from the respective sub-domains. 

Here, bulk data points are first undersampled using the vorticity cut-off based sampling technique previously described. Once obtained, one needs to  keep the temporal resolution for the near-field data points such that, $\Delta t_{Bulk}/ T >= 1/2f_{max}$. On the other hand, the far-field does not have to adhere to such  strict resolution restriction and can be,  $\Delta t/ T < 1/ 2f_{max}.$ This is possible as PINNs are capable of recovering smoother features  even under high temporal sparsity as seen earlier. 

Data points in the near-field were sampled in time with $\Delta t_{Bulk}^{NF} = 0.125$; in the wake, data points were undersampled such that, $\Delta t_{Bulk}^{FF} = 4\times \Delta t_{Bulk}^{NF} = 0.5$.  
This indirectly could give more weightage to the near-field data in time and space. As a result, the velocity flow-field reconstruction and pressure recovery are expected to improve overall. It will be later shown through experiments that increasing temporal sparsity in the near-field can be quite detrimental, especially when the spectral content in the flow becomes richer.

\begin{table}[!htbp]
	\centering
	\caption{ Details of training dataset for vorticity cutoff (VS), and preferential spatio-temporal (PVS) sampling over different domains $\Omega^r_1, \Omega^r_2,$ and $\Omega^r_3$ for the quasi-periodic case considered. The training datasets are generated with  vorticity-based selective sampling ratio $S_{\Omega_z} = 5\%$ with $|\omega_z^*|=0.1$, the   near-field region (NF) being $[-1.5c,1c]$ in the x direction;   $\Delta t/ T_{Bulk}^{NF} = 0.125$; $\Delta t/ T_{Bulk}^{FF} \in {0.125, 0.5}$,  depending on the far-field (FF) spatio-temporal sampling.}
	\resizebox{\textwidth}{!}{\begin{tabular}{ccccccc}
		\hline
		\textbf{Sampling} & Domain & $(\Delta t/T)_{Bulk}^{NF}, (\Delta t/T)_{Bulk}^{FF}$ & $N_{Bulk}$ & $N_{Phy}$& $N_{Bulk}^{NF}/ N_{Bulk}$ (in \%) & $N_{IB}/N_{Bulk}$ (in \%) \\
		\hline
            \multicolumn{7}{c}{Training datasets}\\
            \hline
            VS & $\Omega^r_1$ &  (0.125,0.125) &$1.219e06$ & $2.875e07$ & 14.31 & 52.57\\
            PVS & $\Omega^r_1$ & (0.125,0.5) & $4.756e05$& $2.875e07$ & 36.67 & 134.76 \\
            VS & $\Omega^r_2$ & (0.125,0.125) &$5.805e05$ &$1.808e07$& 30.05 & 110.42 \\
            PVS & $\Omega^r_2$ & (0.125,0.5) & $2.904e05$ & $1.808e07$ & 60.05 & 220.78 \\
            VS & $\Omega^r_3$ & (0.125,0.125) & $2.148e05$ &  $1.077e07$ & 100.0 & 298.40 \\
            \hline
        \end{tabular}}
	\label{tab:prefsample_data-sets}
 \end{table}

\begin{figure}[!htbp]
    \centering
    \includegraphics[clip=True, trim={0 6.5cm 0 0}, width=\linewidth]{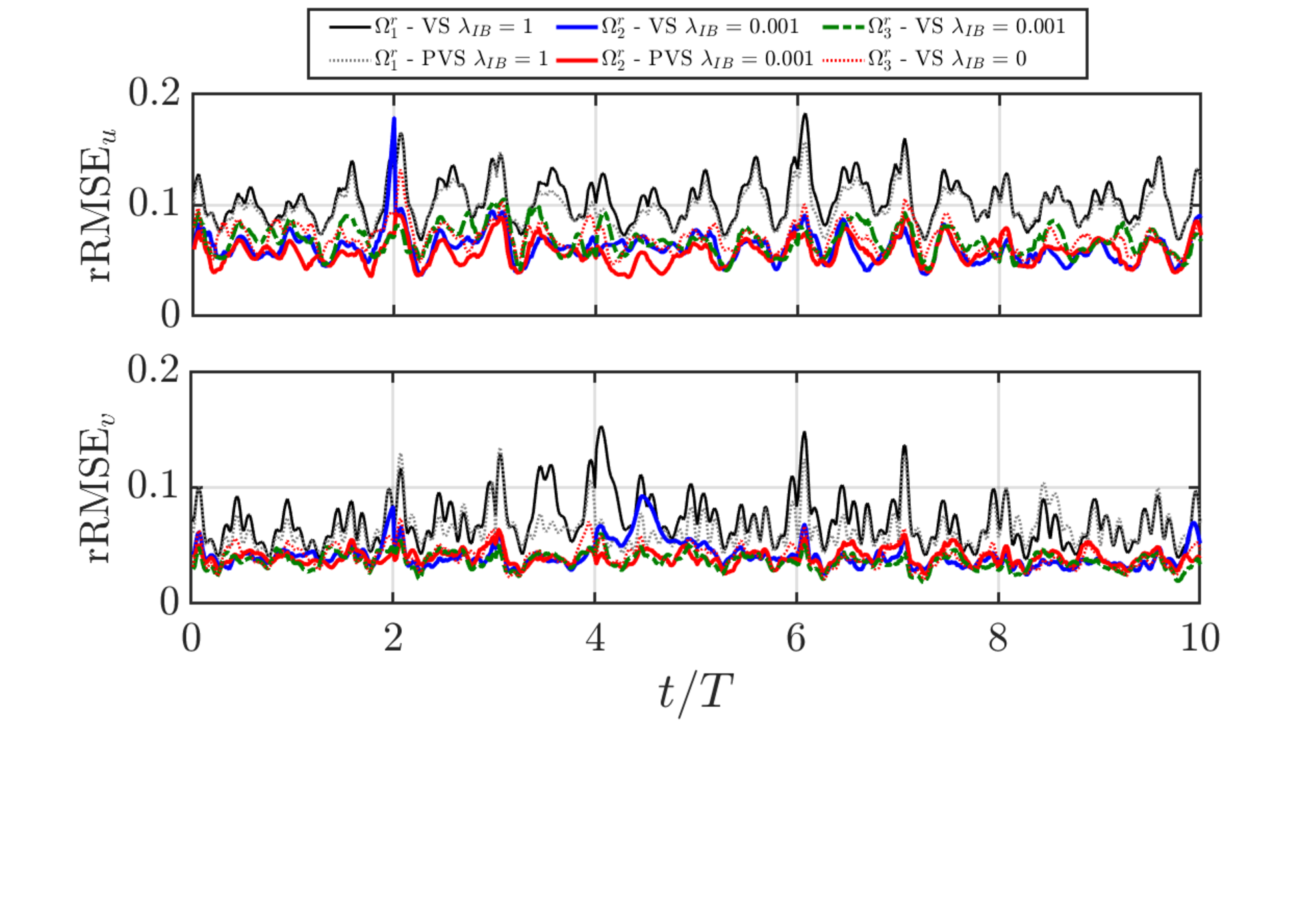}
    \caption{Comparison of relative errors in time for velocity reconstruction for $\mathbf{M2}_{TD+TLLR}$ models trained with and without preferential sampling.}
    \label{fig:relerr_prefsampling}
\end{figure}

To demonstrate the efficacy of PVS, three different domains and datasets have been  considered, $\Omega^r_1: [-1.5c,6.5c]\times[-2c,2c]$, $\Omega^r_2: [-1.5c,3.5c]\times[-2c,2c] $, and $\Omega^r_3: [-1.5c,1.5c]\times[-2c,2c]$. Here, the domain size is decreased to study the importance of the near-field over wake region. Figure~\ref{fig:prefsampling} shows the schematic, and table~\ref{tab:prefsample_data-sets} presents more details of the datasets. Note that, $\Omega^r_1$ was earlier chosen so as to keep at least two vortex couples within the extent. Whereas, the horizontal (x-axis)  extent of $\Omega^r_2$ is smaller and same as that of the periodic  case study but captures only a single vortex couple;  $\Omega^r_3$ is only  the near-field region encompassing the moving body. The wake/far-field regions for $\Omega^r_1$ and $\Omega^r_2$ domains are, $\Omega^r_1 - \Omega^r_3,$ and $\Omega^r_2 - \Omega^r_3,$ respectively. 

\begin{figure}[!htbp]
    \centering
    \includegraphics[width=0.9\linewidth]{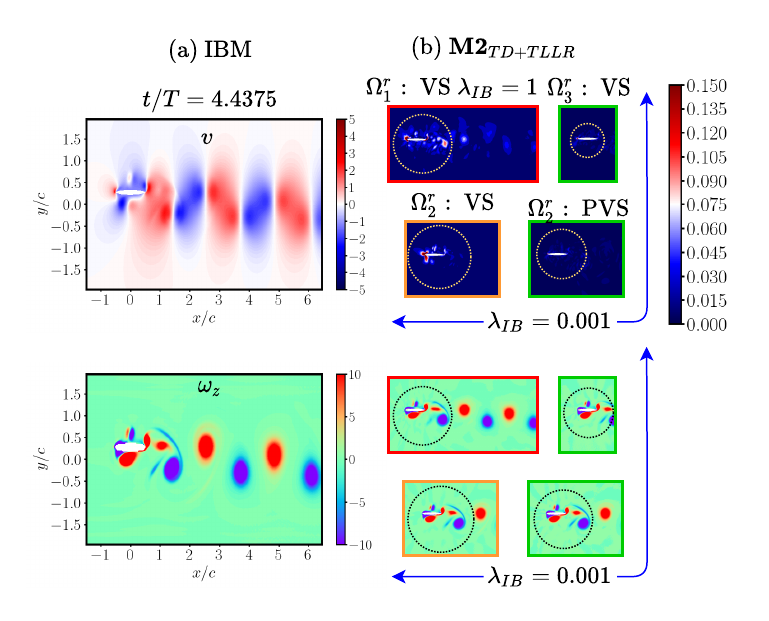}
    \caption{Comparison of, (a) ground truth IBM of velocity component ($v$), and, (b) point-wise errors in model reconstruction at a test time stamp of  $t/T = 4.4375$. Corresponding vorticity snapshots are in the bottom row. }
    \label{fig:prefcontours}
\end{figure}

In addition to a relaxed $\mathcal{L}_{Phy}$, weight coefficient $\lambda_{IB}$ of the $\mathcal{L}_{IB}$ was also relaxed. This was to ease the training by relaxing the competition between the three objectives, $\mathcal{L}_{IB}$ on the moving boundary,  and $\mathcal{L}_{Bulk}, \mathcal{L}_{Phy}$ near the body.
Relative velocity reconstruction errors in figure~\ref{fig:relerr_prefsampling} and the average error metrics in table~\ref{tab:detailedaccuracy_pvs_quasiperiodic} show the remarkable power of combining the preferential spatio-temporal vorticity based sampling (PVS) along with the $\mathcal{L}_{IB}$ loss relaxation. The efficacy of this approach is even more evident from figure~\ref{fig:prefcontours}, where the point-wise velocity reconstruction errors are seen to come down significantly as the domain size reduces along with PVS and $\lambda_{IB}$ relaxation. Note that in the case of $\Omega^r_3,$ the near-field region alone suffices for good velocity reconstruction as indicated by the vorticity contours.

\begin{table}[!htbp]
\centering
\caption{Comparison of the accuracy  of  velocity component ($v$) reconstructed by the standard MB-PINN and variants of seq-MB-PINN when trained with varying levels of temporal sparsity over 10 plunging cycles (Quasi-periodic case)}
\label{tab:detailedaccuracy_pvs_quasiperiodic}
\resizebox{\textwidth}{!}{\begin{tabular}{cccccccc}
	\hline 
	Domain & $\lambda_{IB}$ & Sampling&  \textbf{Network size} & \textbf{RMSE} & \textbf{MAE} & $R^2$ & \textbf{rRMSE}(in \%)\\ 
        \hline 
        \multicolumn{8}{c}{\textbf{$\mathbf{M2}_{TD+TL}$}} \\
        \hline
        $\Omega^r_1$&1&0.5 & $100 \times 5$ & 2.14e-01&	1.03e-01&	9.38e-01&	23.67 \\ 
        $\Omega^r_1$&1&0.125 &$100 \times 10$ &  6.34e-02&	3.13e-02&	9.95e-01&	7.02 \\ 
        $\Omega^r_2$&1&VS &$100 \times 10$ & 5.43e-02&	2.72e-02&	9.96e-01&	6.00\\
        $\Omega^r_3$&1&VS &$100 \times 10$ & 5.8e-01& 2.85e-02& 9.951e-01& 6.58\\
        $\Omega^r_1$&1& PVS & $100 \times 5$ & 5.23e-01 &2.69e-02 &9.964e-01 &5.79  \\ 
        $\Omega^r_2$&1& PVS &$100 \times 10$ & 5.35e-02 & 2.86e-02& 9.963e-01& 5.92\\
        $\Omega^r_2$&0.001& VS &$100 \times 10$ & 3.5e-02& 1.49e-02&9.984e-01 &3.88\\
        $\Omega^r_2$&0.001& PVS &$100 \times 10$ & 3.35e-02& 1.62e-02&9.986e-01 &\textbf{3.7}\\
        \hline
        \multicolumn{8}{c}{$\mathbf{M2}_{TD+TLLR}$ } \\
        \hline
        $\Omega^r_1$&1&0.5 &$100 \times 5$ & 2.21e-01&	1.11e-01&	9.33e-01&	24.34 \\ 
        $\Omega^r_1$&1&0.125 &$100 \times 10$ &  6.49e-02&	3.09e-02&	9.94e-01&	7.18 \\ 
    	$\Omega^r_2$&1&0.125, VS &$100 \times 10$ & 5.33e-02 & 2.55e-02&9.962e-01 &5.9\\
        $\Omega^r_3$&1&0.125, VS &$100 \times 10$ & 5.71e-02& 2.76e-02& 9.952e-01&6.47\\
        $\Omega^r_1$&1& PVS & $100 \times 5$ & 5.86e-02& 3.28e-02& 9.956e-01&6.47  \\ 
        $\Omega^r_2$&1& PVS &$100 \times 10$ & 5.15e-02& 2.5e-02& 9.965e-01& 5.7\\
        $\Omega^r_2$&0.001& VS &$100 \times 10$ & 3.74e-02& 1.57e-02& 9.982e-01& 4.15\\
        $\Omega^r_2$&0.001& PVS &$100 \times 10$ & 3.6e-02 & 1.55e-02&9.984e-01 &\textbf{3.99}\\
        $\Omega^r_3$&0.001& VS &$100 \times 10$ & 3.23e-02& 1.17e-02& 9.986e-01&\textbf{3.66}\\
        $\Omega^r_3$& 0 & VS &$100 \times 10$ & 3.57e-02& 1.25e-02& 9.983e-01&4.05\\
    \hline
\end{tabular}}
\end{table}

The aperiodic flow-fields could differ from solver to solver due to the underlying numerical schemes involved. As a result, the pressure recovered by PINN models trained on IBM data could still demonstrate high point-wise errors, when compared with other solvers (say an ALE solver) as was discussed in~\cite{sundar2024physics}.This is demonstrated in figure~\ref{fig:alepressure_quasiperiodic} of ~\ref{sec:aleibmpressure} where further discussion on this discrepancy is provided. Therefore, it is necessary to determine whether the pressure prediction from PINNs are reasonable. This is possible if one has the time histories of the aerodynamic load coefficients ($C_L$, $C_D$) available for the exact flow-field that was used for training. 
Although pressure is not directly available from the IBM results, the aerodynamic loads can be computed in the IBM routine using the momentum forcing term  and temporal derivative terms of the momentum conservation equations~\cite{majumdar2020capturing}. Since computation of $C_L$ directly depends on the pressure distribution on the moving body, it can serve as a good  metric to indirectly determine the accuracy of the  pressure recovery. In the rest of the discussion, $C_L$ obtained from IBM and reconstructed from the PINN model have been quantitatively compared to evaluate the performance of pressure recovery. To further validate the load time histories qualitatively, $C_L-C_D$ phase portraits have been considered as well.

\begin{table}[!htbp]
\centering
\caption{Comparison of accuracy of loads reconstructed by the seq-MB-PINN (TL and TLLR) for the quasi-periodic case, The models are trained with varying $\lambda_{IB}$ at a fixed $\Delta t_{Bulk}/T = 0.125 (1/8)$) satisfying the Nyquist criteria ($\Delta t/T\leq 1/(2f_{max}) = 0.25$). The sub-networks are of depth, $l = 10,$ and width, $n = 100$ trained over 5 segments of 2 plunging  periods each. Here, true $\hat{C}_L^{Peak} = 14.043$}
\label{tab:detailedaccuracy_loads}
\resizebox{\textwidth}{!}{\begin{tabular}{ccccccc}
	\hline 
	Domain & $\lambda_{IB}$ &  Sampling & $C_L^{Peak}$ &  $\frac{C_L^{Peak} - \hat{C}_L^{Peak}}{\hat{C}_L^{Peak}}$ & $\frac{\textbf{MAE}}{|\hat{C}_L|_{L_{\infty}}}$ (in \%) & $\frac{\textbf{MAE}}{|\hat{C}_L|_{L_{1}}} $ (in \%)  \\ 
        \hline
        \multicolumn{7}{c}{ TL} \\
        \hline
        $\Omega^r_1$ & 1 & 0.5, VS & 5.98  & 59.89 & 52.17 & 76.13 \\
        $\Omega^r_1$ & 1 & VS &10.91  & 22.32 & 25.04 & 36.54 \\
        $\Omega^r_2$ &1 &VS & 11.40  & 18.80 & 17.30 & 25.25 \\ 
        $\Omega^r_3$ &1 &VS & 12.21  & 13.04 & 13.58 & 19.83 \\ 
        $\Omega^r_1$ & 1 & PVS &10.65  & 25.54 & 22.07 & 32.21 \\
        $\Omega^r_2$ & 1 & PVS & 12.27  & 12.61 & 15.59 & 22.76 \\
        $\Omega^r_2$& 0.001 &VS& 13.28 &	5.41 & 5.54 & 8.086 \\ 
        $\Omega^r_2$&\textbf{ 0.001 }&PVS& \textbf{13.37 }&	\textbf{4.80} & \textbf{4.48} &	\textbf{6.54}\\ 
        
    \hline
    \multicolumn{7}{c}{TLLR} \\
    \hline
        $\Omega^r_1$ & 1 & 0.5, VS & 8.19  & 41.67 & 46.23 & 67.47 \\
        $\Omega^r_1$ & 1 & VS & 9.83  & 30.50 & 25.66 & 37.46 \\
        $\Omega^r_2$ &1 &VS  & 11.64 & 17.12 & 17.80 & 25.25 \\ 
        $\Omega^r_3$ &1 &VS  & 10.44 & 25.59 & 18.91 & 22.76 \\ 
        $\Omega^r_1$ & 1 & PVS & 11.95  & 14.87 & 18.79 & 27.42 \\
        $\Omega^r_2$ & 1 & PVS & 12.31 & 12.18 & 15.46 & 22.56 \\ 
        $\Omega^r_2$ &0.001 &VS&  13.33 &	5.01 & 5.33 & 7.77 \\
        $\Omega^r_2$ &0.001 &PVS&  13.41 & 4.52 & 3.81 & 5.56 \\ 
        $\Omega^r_3$ &0.001 &VS& 13.60 &	3.14 & 3.44 & 5.03\\
        $\Omega^r_3$ &0.0 &VS& 13.63 & 2.94 & 3.09 &	4.51 \\ 
    \hline
\end{tabular} }
\end{table} 
\begin{figure}[!htbp]
    \centering
    \includegraphics[width=0.9\linewidth]{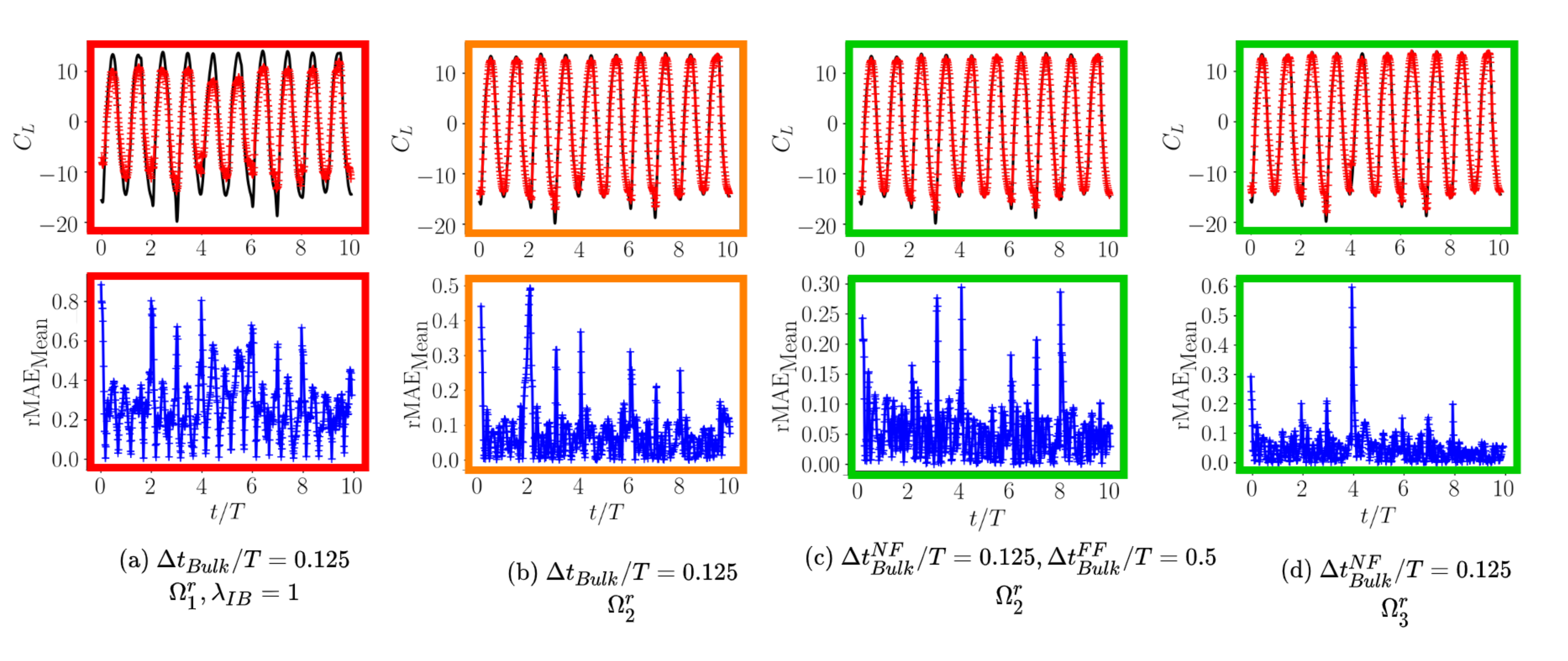}
    \caption{Comparison of $C_L$ reconstructions obtained from $\mathbf{M0}$ and seq-MB-PINN(TLLR) for different bulk data sampling approaches and $\mathcal{L}_{IB}$ weighting.}
    \label{fig:qploads}
\end{figure}

\begin{figure}[!htbp]
    \centering
    \includegraphics[width=\linewidth]{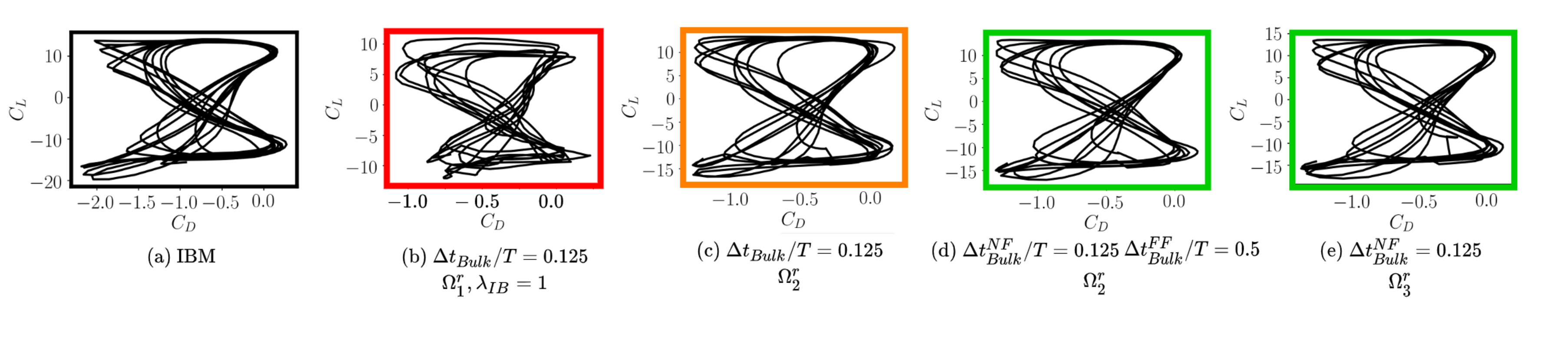}
    \caption{Qualitative of comparison of $C_L-C_D$ phase portraits reconstructed from (a) IBM ground truth data, and (b-e) predictions of $\mathbf{M2}_{TD+TLLR}$ models across different bulk data sampling approaches and $\mathcal{L}_{IB}$ values.}
    \label{fig:clcdphaseportrait_quasiperiodic}
\end{figure}

Note in  table~\ref{tab:detailedaccuracy_loads} that, improving the structure of the time series comes at the cost of under prediction of the $C_L^{Peak}.$ Further training/fine tuning of the models can improve this. But notably, the errors have certainly dropped by relaxing the $\mathcal{L}_{IB}.$ The effect of $\lambda_{IB} = 0$ is such that the errors in load reconstruction become higher in the first subdomain which drop over subsequent domains (see figure~\ref{fig:qploads}). With a slight non-zero weighting, such as  $\lambda_{IB} = 0.001$, the overall errors in all segments remain low.  Qualitatively, the $C_L-C_D$ phase-portraits in figure~\ref{fig:clcdphaseportrait_quasiperiodic} also show a distinct improvement when $\mathcal{L}_{IB}$ relaxation is coupled with preferential spatio-temporal sampling. This deduction is especially useful in the scenario when one needs to train MB-PINNs without any information of the moving body velocity. In this case, one would have to discard the predictions in the first subdomain and consider the results from the subsequent subdomains where transfer learning has come into effect.
However, if one has the moving body velocity information {\it a priori}, then enforcing it through $\mathcal{L}_{IB}$ with a reasonably low $\lambda_{IB}$ proves to be beneficial in load reconstruction over the entire time domain considered. 

Overall, under a fixed training budget, Class II models, especially, sequential learning using time domain decomposition with transfer learning are performant in dealing with different temporal domain complexities. Moreover, it is beneficial to combine preferential spatio-temporal sampling, and relaxation of $\mathcal{L}_{IB}$ loss component, in addition to physics loss relaxation significantly, to improve the overall pressure recovery and load reconstruction for the quasi-periodic case. 

\section{Summary and Conclusions}\label{sec:conclusion}
The present study highlights the limitations of traditional PINNs for long-time integration of unsteady flow and flow-structure interaction problems and demonstrates the benefits of decomposition-based strategies for addressing error accumulation, computational cost, and complex dynamics. The earlier developed moving boundary-enabled PINN (MB-PINN) formulation~\cite{sundar2024physics} under the  immersed boundary aware (IBA) framework has been utilised  for velocity reconstruction and pressure recovery.

Key difficulties for PINNs in long-time integration include temporal sparsity, long temporal domains and rich spectral content. In order to tackle these temporal domain complexities in the context of moving bodies using MB-PINN, two sequential learning strategies have been  explored: - time marching with gradual increase in time domain size, and time domain decomposition with an overall reduced problem size. While the former strategy struggles with error accumulation over long time domains, the latter approach combined with transfer learning effectively reduces error propagation and computational complexity. For the present problem of incompressible unsteady flows past a flapping airfoil exhibiting periodic and quasi-periodic dynamics, the time decomposition approach with preferential spatio-temporal sampling successfully  improves both accuracy and efficiency for the pressure recovery and aerodynamic load reconstruction. 
However, these methods are not yet directly applicable for strongly aperiodic, such as chaotic flows,  due to the underlying neural networks' spectral bias, also the  flow-field trajectories being vastly different under  small variations in the initial condition. 

When flow-field data is available—even if only sparsely—the time domain should be partitioned into subdomains that span one or two plunging time periods. 
Over-partitioning into smaller subdomains would necessitate training an excessive number of sub-networks, thereby increasing computational expenses and heightening the risk of over-fitting.

In this study the models trained  operated only on a single parametric instance, however, in future, operator learning techniques can be adopted to handle multiple parametric instances, or different input conditions such as gusty inflows. This would be motivating towards  building parametric surrogates for unsteady flow control. Also, the models need to become viable for training on more complex and realistic 3D flows, which requires further research on multiple fronts: the underlying neural network architecture, optimization of collocation points sampling, time-adaptive labeled data sampling, adaptive loss balancing and optimization techniques which can accelerate loss convergence and training. The present study also holds promise for experimentalists, where PIV measurements of planar velocity fields can be utilized to recover hidden quantities and infer the underlying load generation mechanisms.

\pagebreak
\bibliographystyle{elsarticle-num} 
\bibliography{bibliography.bib}

\pagebreak
\appendix
\section{Temporal resolution of residual collocation points}\label{sec:reseffect}
The accuracy details for velocity reconstruction and pressure recovery for baseline $M0$ models trained with different temporal resolution of residual collocation points $\Delta t_{Phy}/T$ are presented in tables~\ref{tab:resaccuracyv} and \ref{tab:resaccuracyp}. Here, the bulk data is sampled at $\Delta t_{Bulk}/T = 0.5.$ It is observed that the velocity reconstruction errors improve significantly up to $\Delta t_{Phy}/T = 0.0125$ beyond which the improvement is not significant. The significant improvement in velocity reconstruction is met with only a slight degradation of pressure recovery. But further increase in the temporal resolution of residual collocation points would make it computationally expensive when training over large time domains.

\begin{table}[!htbp]
\centering
\caption{Comparison of accuracy details of $y$ velocity component ($v$) reconstructed by the standard MB-PINN when trained with a varying number of PDE collocation points in time ($\Delta t_{Phy}/T$). Here, the temporal sparsity of data snapshots is such that $\Delta t_{Bulk}/T = 0.5$ not satisfying the Nyquist criteria ($\Delta t/T\leq 1/(2f_{max}) = 0.25$). The errors are evaluated over two plunging cycles. The network is of depth $l = 10,$ and width $n = 100.$}
\begin{tabular}{ccccc}
	\hline 
	$\Delta t_{Phy}/T$ &  \textbf{RMSE} & \textbf{MAE} & $R^2$ & \textbf{rRMSE}(in \%)\\ 
        \hline 
        \multicolumn{5}{c}{\textbf{Linear Interpolation}} \\
        \hline
        -&5.17e-01&3.03e-01&-1.01e-01&104.96\\
        \hline
        \multicolumn{5}{c}{\textbf{MB-FNN}} \\
        \hline
	-& 3.56e-01&	1.87e-01&	3.77e-01&	69.87 \\ 
 \hline
        \multicolumn{5}{c}{\textbf{$\mathbf{M0}$}} \\
        \hline
        0.5 &  2.18e-01&	1.22e-01&	7.68e-01& 42.67 \\ 
	0.2 & 1.50e-01&	6.51e-02&	8.94e-01& 29.82 \\ 
	0.1 & 1.37e-01&	5.59e-02&	9.11e-01& 27.27 \\ 
        0.05 & 6.46e-02&	2.69e-02&	9.81e-01& 12.72 \\ 
        0.025 & 5.62e-02&	2.38e-02&	\textbf{9.87e-01}& \textbf{11.07}\\ 
        0.0125 & 4.32e-02&	1.95e-02&	\textbf{9.91e-01}& \textbf{8.45} \\ 
        0.00625 & 4.21e-02&	1.94e-02&	\textbf{9.92e-01}& \textbf{8.22} \\ 
    \hline
\end{tabular}
\label{tab:resaccuracyv}
\end{table}

\begin{table}[!htbp]
\centering
\caption{Comparison of accuracy details of pressure ($p$) recovered by the standard MB-PINN when trained with a varying number of PDE collocation points in time ($\Delta t_{Phy}/T$). Here, the temporal sparsity of data snapshots is such that $\Delta t_{Bulk}/T = 0.5$ not satisfying the Nyquist criteria ($\Delta t/T\leq 1/(2f_{max}) = 0.25$). The errors are evaluated over two plunging cycles. The network is of depth $l = 10,$ and width $n = 100.$}
\begin{tabular}{ccccc}
	\hline 
	$\Delta t_{Phy}/T$ &  \textbf{RMSE} & \textbf{MAE} & $R^2$ & \textbf{rRMSE}(in \%)\\ 
        \hline
        0.5 &  1.28 &	8.25e-01&	5.77e-02&	96.72 \\ 
	0.2 & 1.22&	7.32e-01&	1.53e-01&	91.35 \\ 
	0.1 & 9.72e-01&	5.45e-01&	4.43e-01&	73.77 \\ 
        0.05 & 3.79e-01&	2.20e-01&	8.95e-01&	29.75 \\ 
        0.025 & 2.82e-01&	1.77e-01&	\textbf{9.47e-01}&	\textbf{21.74} \\ 
        0.0125 & 2.93e-01&	1.82e-01&	\textbf{9.44e-01}&	\textbf{22.55} \\ 
        0.00625 &2.76e-01&	1.70e-01&	\textbf{9.49e-01}&	\textbf{21.17} \\ 
    \hline
\end{tabular} 
\label{tab:resaccuracyp}
\end{table}

\section{Effect of sparsity on pressure recovery}\label{sec:sparsityeffect}

Here in table~\ref{tab:detailedaccuracy_pressurerecovery_stdpinn}, the accuracy details for pressure recovery using \textbf{M0} models of different sizes across different temporal sparsity levels is presented. Clearly, a long as the temporal resolution satisfies Nyquist criteria, the pressure recovery errors are lover than 15\% for the 100 x 10 network. Importantly, it was also observed that in the case of a forward problem ($\Delta t_{Bulk}/T = \infty,$ with velocity data vailable only at the initial time stamp $t/T = 0$), the pressure recovery drastically suffers even over a time domain of two plunging time periods. Forward problems involving moving boundaries need special attention in terms of initialisation, training, architecture, loss formulation, and architecture. This requires an in depth independent investigation which is out of the scope of the present work. 
\begin{table}[!htbp]
\centering
\caption{Comparison of pressure recovery accuracy details of the standard MB-PINN ($\mathbf{M0}$) model when trained with varying levels of temporal sparsity. Here, in the table (F) stands for forward problem when $\Delta t_{Bulk}/T = \infty$}
\label{tab:detailedaccuracy_pressurerecovery_stdpinn}
\resizebox{\textwidth}{!}{\begin{tabular}{cccccc}
	\hline 
	$\Delta t_{Bulk}/T$ &  \textbf{Network size} &\textbf{RMSE} & \textbf{MAE} & $R^2$ & \textbf{rRMSE}(in \%) \\ 
	\hline 
        0.2 &$100 \times 10$ & 1.372e-01 & 8.16e-02 & 9.861e-01 & 10.84 \\ 
	0.25 &$100 \times 10$ & 1.78e-01&	1.07e-01&	9.79e-01&	13.86 \\
	0.3 &$100 \times 10$ &  2.22e-01&	1.25e-01&	9.68e-01&	16.99\\ 
	0.4 &$100 \times 10$ & 2.31e-01&	1.48e-01&	9.62e-01&	18.05 \\ 
	0.5 &$100 \times 10$ & 2.82e-01&	1.77e-01&	9.47e-01&	21.74 \\ 
     $\infty$(F) &$100 \times 10$ & 9.46e-01&	6.84e-01&	4.32e-01&	73.70 \\
	\hline 
        0.2 &$100 \times 5$ & 2.88e-01&	1.85e-01&	9.48e-01&	22.12\\ 
	0.25 &$100 \times 5$ & 3.02e-01&	1.90e-01&	9.43e-01&	23.4 \\ 
	0.3 &$100 \times 5$ & 4.19e-01&	2.73e-01&	8.69e-01&	32.66 \\ 
	0.4 &$100 \times 5$ & 4.98e-01&	3.27e-01&	8.36e-01&	38.32 \\ 
	0.5 &$100 \times 5$ & 7.22e-01&	4.87e-01&	6.95e-01&	54.71 \\ 
  $\infty$(F) &$100 \times 5$ & 1.39&	1.06&	-1.85e-01&	107.49 \\
	\hline 
       
        0.2 &$50 \times 5$ & 5.25e-01&	3.66e-01&	8.38e-01&	39.79 \\ 
	0.25 &$50 \times 5$ & 5.66e-01&	3.93e-01&	8.09e-01&	43.15 \\ 
	0.3 &$50 \times 5$ & 6.99e-01&	4.87e-01&	6.93e-01&	53.45 \\ 
	0.4 &$50 \times 5$ & 8.56e-01&	5.97e-01&	5.59e-01&	64.84 \\ 
	0.5 &$50 \times 5$ & 1.05&	7.41e-01&	3.72e-01&	79.07 \\ 
  $\infty$(F) &$50 \times 5$ & 1.66 &	1.31 &	-6.66e-01&	127.04 \\
    \hline
\end{tabular} }
\end{table}

\section{Comparison with ALE pressure data}\label{sec:aleibmpressure}
The pressure predictions of $\mathbf{M2}_{TD+TLLR}$ models trained with and without preferential sampling were compared with ALE pressure data in the same spirit as in section~\ref{sec:periodicresults}.
\begin{figure}[!htbp]
    \centering
    \includegraphics[width=\linewidth]{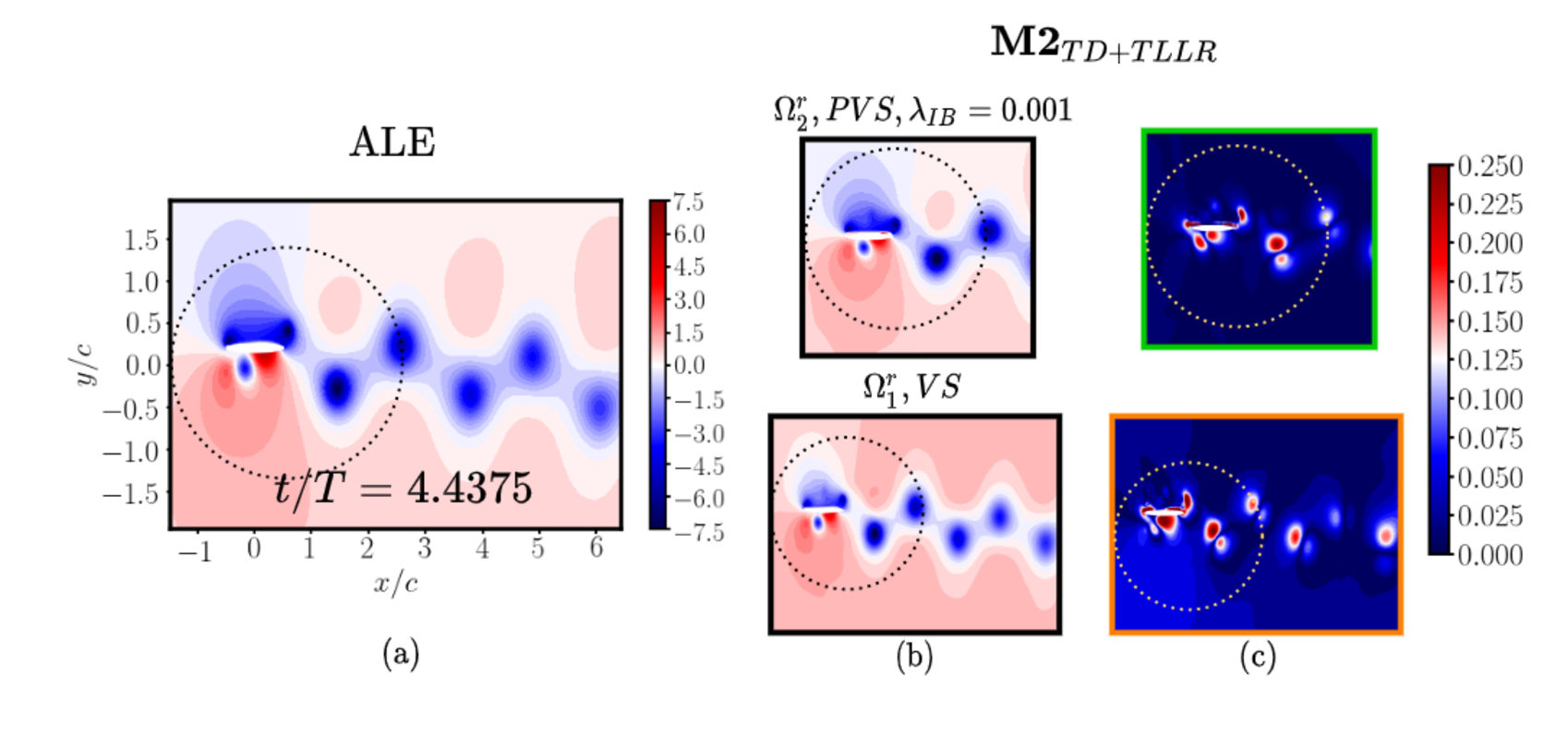}
    \caption{Comparison of pressure contours of (a) ALE and (b) $\mathbf{M2}_{TD+TLLR}$ model predictions at a representative test time stamp $t/T = 4.4375$, and (c) corresponding pointwise maximum-value normalized absolute error contours. In (b-c) Top row corresponds to the best $\mathbf{M2}_{TD+TLLR}$ model trained on $\Omega^r_2$ domain with preferential sampling and $\lambda_{IB} = 0.001.$ In the bottom row, the predictions correspond to $\mathbf{M2}_{TD+TLLR}$ model trained on $\Omega^r_1$ domain with just vorticity cutoff sampling and $\lambda_{IB} = 1.$}
    \label{fig:alepressure_quasiperiodic}
\end{figure}
Due to flow aperiodicity, and the underlying numerical schemes being different in ALE and IBM solvers, the predictions of models trained on IBM velocity data do not completely match that of ALE. This is observed from the high pointwise errors wherever vortices are present in the flow (see figure~\ref{fig:alepressure_quasiperiodic}) even for the best model obtained using preferential sampling and $\mathcal{L}_{IB}$ weighting.

\end{document}